\documentclass[twocolumn,english,aps,pra,reprint, superscriptaddress,showpacs,longbibliography,showkeys]{revtex4-2}

\usepackage{amsmath,amssymb,graphicx,color,xcolor,braket}
\usepackage[colorlinks,citecolor=blue,urlcolor=blue,linkcolor = blue]{hyperref}
\usepackage{physics}
\usepackage{tikz, ifthen}
\usetikzlibrary{arrows.meta}
\usetikzlibrary{decorations.pathreplacing,calc}
\usepackage{float}
\makeatletter
\let\newfloat\newfloat@ltx
\makeatother
\usepackage{orcidlink}
\usepackage{quantikz}
\usepackage{placeins}
\usepackage{longtable}

\makeatletter\renewcommand{\p@subsection}{}\makeatother

\makeatletter
\AtBeginDocument{%
  \renewcommand\section{\@startsection{section}{1}{\z@}%
    {0.8cm \@plus1ex \@minus .2ex}{0.5cm}%
    {\normalfont\small\bfseries\raggedright}}%
  \renewcommand\subsection{\@startsection{subsection}{2}{\z@}%
    {.8cm \@plus1ex \@minus .2ex}{.5cm}%
    {\normalfont\small\bfseries\raggedright}}%
}
\makeatother

\graphicspath{{Figures/}}

\makeatletter
\newcommand{\equalcontrib}{%
  \frontmatter@footnote{These authors contributed equally to this work.}}
\makeatother

\definecolor{rampcool}{HTML}{2166AC}
\definecolor{rampwarm}{HTML}{B2182B}
\definecolor{oigrey}{HTML}{999999}
\definecolor{oidarkgrey}{HTML}{4D4D4D}

\colorlet{prob_colour}{rampcool}
\colorlet{driver_colour}{rampwarm}
\colorlet{prep_colour}{oigrey}
\colorlet{corr_colour}{oidarkgrey}
\colorlet{repeat_colour}{oidarkgrey}

\newcommand{\Sthree}{\ensuremath{S^{3}}}

\begin{document}

\author{Merle Stahl~\orcidlink{0009-0005-8652-2356}\equalcontrib}
\affiliation{Institute for Computational Systems Biomedicine, University of Hamburg, 22761 Hamburg, Germany}

\author{Robert~J.~Banks~\orcidlink{0009-0004-5198-1651}\equalcontrib}
\email[Corresponding author: ]{r.banks@parityqc.com}
\affiliation{Parity Quantum Computing Germany GmbH, 20095 Hamburg, Germany}

\author{Matthias~Traube~\orcidlink{0000-0002-4229-6829}}
\affiliation{Parity Quantum Computing Germany GmbH, 20095 Hamburg, Germany}

\author{Josua~Unger~\orcidlink{0000-0003-1633-1843}}
\affiliation{Parity Quantum Computing GmbH, A-6020 Innsbruck, Austria}

\author{Wolfgang~Lechner~\orcidlink{0000-0003-3662-1020}}
\affiliation{Parity Quantum Computing Germany GmbH, 20095 Hamburg, Germany}
\affiliation{Parity Quantum Computing GmbH, A-6020 Innsbruck, Austria}
\affiliation{Institute for Theoretical Physics, University of Innsbruck, A-6020 Innsbruck, Austria}

\author{Jan Baumbach~
\orcidlink{0000-0002-0282-0462}}
\affiliation{Institute for Computational Systems Biomedicine, University of Hamburg, 22761 Hamburg, Germany}
\affiliation{Department of Mathematics and Computer Science, University of Southern Denmark, 5230 Odense, Denmark}

\author{Mhaned Oubounyt~ 
\orcidlink{0000-0001-5011-6683}}
\affiliation{Institute for Computational Systems Biomedicine, University of Hamburg, 22761 Hamburg, Germany}

\begin{abstract}
Protein–protein interaction (PPI) network alignment combines topological and sequence information to identify conserved modules across species, but global alignment remains challenging: heuristics sacrifice optimality, while exact methods lack scalability.
We model the alignment as a weighted maximum common induced subgraph problem and reformulate it through the modular product graph to a minimum-weight vertex cover on the complement, with node weights carrying sequence similarity. To solve this problem, we develop a hybrid framework combining kernelisation, branch-and-bound, and seven Quantum Approximate Optimisation Algorithm (QAOA) formulations. These formulations differ in how the cover constraints are enforced, from penalty terms in the cost Hamiltonian to mixers confined to the feasible subspace. For single round QAOA, we derive closed-form expressions for the expected cost of four circulant mixer variants, enabling performance characterisation without circuit simulation.
Applied to synthetic and real-world networks reduced to KEGG pathways, the QAOA formulations achieve high topological conservation on the aligned core while at least maintaining biological conservation comparable to leading classical aligners, at the cost of reduced node coverage. Across selected KEGG pathways, the aligned subnetworks retain disease-associated proteins, preserving biologically relevant information. Cheaper formulations leave more edges uncovered, while enforcing feasibility in the mixer raises circuit depth by one to two orders of magnitude.
Together, these results highlight the potential of quantum optimisation for PPI network alignment and the resource trade-offs that will shape its scalability as quantum hardware matures.

\begin{figure}[H]
    \centering
\includegraphics[width=0.8\linewidth]{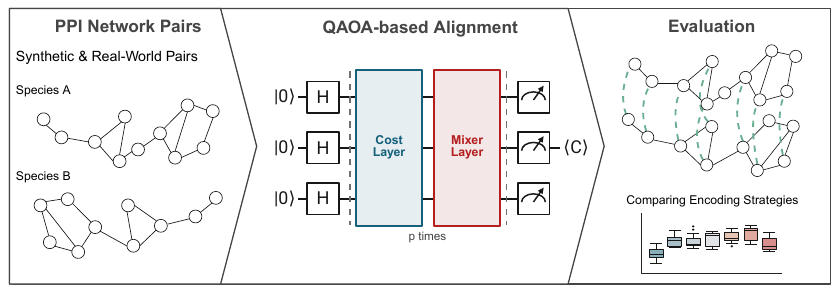}
\end{figure}
\end{abstract}


\date{4 September 2026}

\title{Quantum Optimisation for Protein-Protein Interaction Network Alignment}

\maketitle

\section{Introduction}
\label{sec:intro}

Protein-protein interaction (PPI) networks map the molecular machinery of the
cell, with nodes representing proteins and edges to their physical or functional
associations. Because a protein's function is largely determined by its
interaction partners, aligning PPI networks across species allows conserved
functional modules to be identified and functional annotation to be transferred
between organisms, including from model organisms to humans
\cite{maliao2020}. This matters for translational research: much of what is known about human gene
function and disease biology was first established in model organisms such as mice, and
network alignment offers a systematic way to transfer that knowledge across species.

Formally, global network alignment seeks an injective mapping between the
proteins of two networks that maximises the number of conserved interactions,
or equivalently the maximum common subgraph they share~\cite{singh2008}. This is a maximum
clique in their modular product graph \cite{levi1973,barrow1976}, or a minimum
vertex cover in its complement \cite{karp1972}, and the problem is NP-hard.

Existing aligners resolve the resulting intractability in one of two ways, and
neither is fully satisfactory. Heuristic methods such as IsoRank
\cite{singh2008}, MAGNA++ \cite{vijayan2015}, HubAlign \cite{hashemifar2014},
L-GRAAL \cite{maloddognin2015}, SANA \cite{mamano2017}, and AntNetAlign
\cite{rodriguez2023} scale to whole interactomes but can return suboptimal
solutions. Exact methods, including the Lagrangian-relaxation aligner Natalie2
\cite{elkebir2015} and dedicated branch-and-bound MCIS solvers
\cite{mccreesh2017,zhou2022,yu2025}, do certify optimality and remain efficient
on sparse instances, but they slow down quickly as the modular product graph
gets denser. Cross-species PPI alignment generates exactly this regime: sequence
homology is permissive, so many protein pairs are admissible candidates. No
existing method offers both guaranteed optimality and tractable scale, which
motivates the search for alternative solvers.

We investigate whether quantum optimisation can close this gap. The Quantum Approximate
Optimisation Algorithm (QAOA) encodes a combinatorial problem as a cost
Hamiltonian whose ground state corresponds to the optimal solution and whose
low-lying excited states correspond to near-optimal ones, and alternates
evolution under this Hamiltonian with a mixing Hamiltonian that drives
transitions between candidate solutions \cite{farhi2014}. 
For constrained problems such as MCIS, feasibility can be enforced softly,
through penalty terms in the cost Hamiltonian, or structurally, by restricting
the mixer to the feasible subspace
\cite{hadfield2019,wangrubin2020,baertschi2020}.
The two routes differ in qubit
count, circuit depth, and whether the sampler returns feasible solutions.

No existing quantum work compares them on the PPI network alignment problem.
Daskin, Grama and Kais~\cite{daskin2014} with a follow-up by Daskin~\cite{daskin2015} align PPI
networks by quantum phase estimation on the Kronecker-product transition
matrix, an eigenvector method rather than a combinatorial one. Huang and
Roje~\cite{huang2021} cast MCIS isomorphism as a quadratic
unconstrained binary optimisation (QUBO) for quantum annealing, and
Prasad~\cite{prasad2025} embeds QAOA-derived bounds in a classical
branch-and-bound for maximum clique. Both works address generic graphs without biological
application, and neither compares constraint-handling strategies. 

The general standing of QAOA is mixed. At low depth on high-girth regular
graphs, local classical algorithms
match or exceed QAOA on Max-Cut \cite{marwaha2021}, and barren plateaus impede
trainability at scale \cite{mcclean2018,wangfontana2021}. The picture at higher
depth is more favourable. On large-girth $D$-regular graphs the $p=11$ QAOA
exceeds all classical algorithms known to the authors that rest on no unproven
conjectures \cite{basso2022,farhi2025}, and on the Low Autocorrelation Binary Sequence problem QAOA at constant
depth scales better than state-of-the-art branch-and-bound solvers in simulations
up to 40 qubits, with the best reported
scaling coming from QAOA inside quantum minimum finding, a form of amplitude
amplification \cite{shaydulin2024}. On hardware, warm-started QAOA exceeds the
cuts returned by Goemans--Williamson on 96-node trapped-ion instances that
classical solvers still solve exactly \cite{he2026}. The regimes where an
advantage would appear lie beyond current devices, and resource estimates place
the crossover in the fault-tolerant regime \cite{guerreschi2019,omanakuttan2025}.
Our focus is therefore on the formulation of QAOA for PPI alignment and its
resource requirements.

We formulate PPI network alignment as a minimum weighted vertex cover on the complement of the
modular product graph, with node weights encoding sequence similarity, and
instantiate seven QAOA variants spanning penalty-based through fully
mixer-constrained feasibility handling. Because direct QAOA is limited by
available qubit count, we embed each variant in a branch-and-bound
decomposition that reduces each instance to smaller subproblems. We evaluate the pipeline on NAPAbench2 synthetic networks \cite{woo2020} and
real human--mouse KEGG pathway subnetworks \cite{kanehisa2000}, benchmarking
against established classical aligners. We contribute a QAOA-amenable formulation of PPI network alignment, a
systematic comparison of how constraint handling trades feasibility against
qubit count and circuit depth, and an assessment of where the pipeline stands
against classical solvers.

\section{Materials and methods}
\label{sec:methods}

\begin{figure*}[t]
    \centering
    \includegraphics[width=\linewidth]{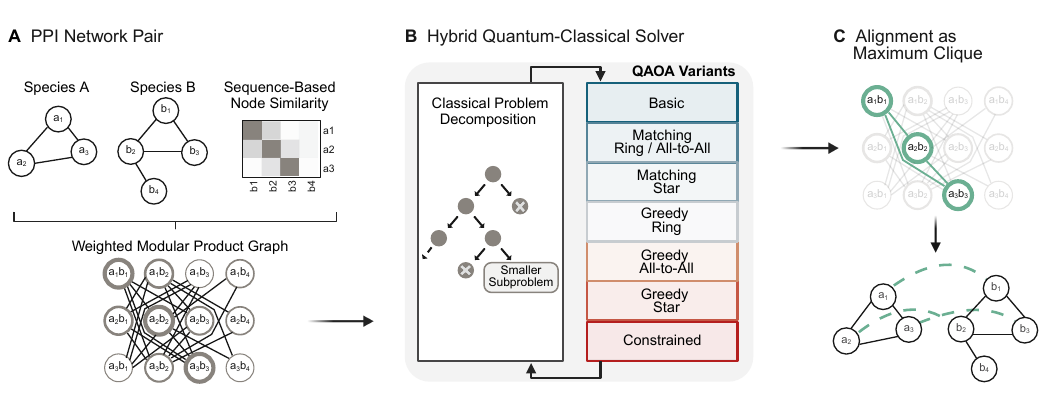}
    \caption{\textbf{Overview of the QAOA-based pipeline for PPI network alignment.} (A) The two PPI networks to be aligned, together with a sequence-based node-similarity matrix between their proteins, are combined into a weighted modular product graph. Its nodes are candidate protein pairs, weighted by sequence similarity, and two nodes are adjacent if the pairs they represent are compatible, i.e.\ if the corresponding proteins are connected in both networks or in neither. (B) The maximum-weight clique problem on the modular product graph is reformulated as a minimum-weight vertex cover problem on the complement graph and reduced to a QUBO. Branch-and-bound with kernelisation shrinks each instance to a qubit-feasible core, which is solved by QAOA in one of seven variants. (C) Sampling, decoding, and post-processing recover the solution, and the resulting maximum-weight clique corresponds to the final optimal node-to-node alignment.}
    \label{fig:pipeline}
\end{figure*}

We develop a hybrid quantum--classical framework for aligning two PPI networks using their modular product graph. The quantum solver is applied to the vertex-cover problem remaining after classical preprocessing (Fig.~\ref{fig:pipeline}).

\subsection{PPI network alignment as weighted maximum common induced subgraph problem}
\label{sec:mcs}

We formulate the alignment of two PPI networks as a weighted maximum common induced subgraph (MCIS) problem. Let $G_1 = (V_1, E_1)$ and $G_2 = (V_2, E_2)$ be the
two PPI networks, each from a different species, where nodes represent proteins
and edges represent interactions. A global network alignment is an injective
partial node mapping $f : V_1 \to V_2$, mapping each protein of the smaller
network with at most one protein of the larger~\cite{maliao2020,faisal2015}.

The search for an optimal such mapping can be cast as a clique-search problem~\cite{klau2009}
through the modular product graph construction. We build a graph whose nodes
are candidate protein pairs $(u, v)$ with $u \in V_1$ and $v \in V_2$,
restricted to biologically plausible pairings~\cite{elkebir2015,clark2014}: pairs with zero weight are dropped before the product graph is
built. This leaves the weighted optimum unchanged, since such a pair
contributes nothing to the objective.
Two nodes $(u, v)$ and $(u', v')$ are connected by an edge if and only if the
pairings are mutually consistent: $u \neq u'$, $v \neq v'$, and the two pairs preserve the same relationship in both
networks, that is, $\{u, u'\} \in E_1 \Leftrightarrow \{v, v'\} \in E_2$. Under
this construction, a set of mutually consistent pairings forms a clique, and a
maximum common subgraph of $G_1$ and $G_2$ corresponds exactly to a maximum
clique in the modular product graph $H$, whose complement we denote
$\bar{H}$~\cite{levi1973,barrow1976}.

Maximum weight clique and minimum weight vertex cover are equivalent problems,
and we work with the vertex cover because it is the form the mixers of
Sec.~\ref{sec:cliquecover} are built on: a clique in $\bar{H}$ forces all but
one of its nodes into any valid cover. A maximum clique in $H$ is a
maximum independent set in $\bar{H}$, and the complement of a
maximum independent set is a minimum vertex cover of $\bar{H}$. The same holds
with weights, since the total node weight is fixed. Solving for a minimum weight
vertex cover on $\bar{H}$ therefore gives the
maximum clique, and hence the optimal alignment \cite{karp1972,lucas2014}.

To incorporate sequence homology, each candidate-pair node carries a weight
derived from the pairwise sequence similarity of the two proteins, normalised
to a common scale, so that pairings between homologous proteins are
preferentially retained in the alignment. Topological conservation is enforced structurally by
the edges of the modular product graph, so the weights determine which of the
topologically consistent alignments is selected, yielding alignments that are
consistent with both network topology and protein homology.

\subsection{Binary formulation: minimum weighted vertex cover}
\label{sec:binary}
 
Let $\bar{H} = (V, E)$ denote the complement of the modular product graph, with
node weights $w_i \geq 0$ carrying the normalised sequence similarity of the
candidate pair $i \in V$. Introducing a binary variable $x_i \in \{0,1\}$ for
each node, with $x_i = 1$ if node $i$ is placed in the vertex cover, minimum
weighted vertex cover on $\bar{H}$ can be written as the QUBO
\begin{equation}
    \label{eq:q_mvc}
    \mathcal{Q}_\text{MVC}(\boldsymbol{x}) = \sum_{i \in V} w_i x_i + \lambda \sum_{(i,j) \in E} (1-x_i)(1-x_j),
\end{equation}
where the second sum penalises every edge whose two endpoints are both left out
of the cover.
 
The penalty satisfies $\lambda\geq \max_{i\in V} w_i$. Since every violating
edge in a vertex cover can be removed by adding one of the nodes in the
violating edge to the vertex cover, and since the cost of this cannot exceed
$\max_{i\in V} w_i$, $\mathcal{Q}_\text{MVC}(\boldsymbol{x})$ can be considered
an upper bound on the minimum vertex cover achieved by $\boldsymbol{x}$ with
simple post-processing. We apply that repair step to the sampled bitstrings in
the branch-and-bound runs of Sec.~\ref{sec:bnb}. The direct-QAOA results are
reported as raw sampler output. Since any vertex cover can be decoded as a
common subgraph, $\mathcal{Q}_\text{MVC}(\boldsymbol{x})$ also acts as a lower
bound on the maximum weighted common subgraph. The QUBO maps straightforwardly
onto an Ising Hamiltonian,
\begin{equation}
    \label{eq:h_mvc}
    H_\text{MVC} = \sum_{i \in V} w_i \frac{1-Z_i}{2} + \lambda \sum_{(i,j) \in E} \frac{1+Z_i}{2}\frac{1+Z_j}{2},
\end{equation}
with one qubit per candidate-pair node and $Z_i$ the Pauli-$Z$ operator on
qubit $i$. Equation~\eqref{eq:h_mvc} is the problem Hamiltonian used throughout
the remainder of this work.
 
In all experiments we set $\lambda = \max_{i\in V} w_i + 1$, which strictly exceeds the bound above.  In practice we re-normalise the rest of the problem Hamiltonian such that $\lambda =1$. Holding $\lambda$ close to its minimum admissible
value keeps the penalty scale comparable to that of the objective.

\subsection{Solving the vertex-cover problem with QAOA}
\label{sec:qaoa}

\paragraph{The quantum approximate optimisation algorithm.}
 
QAOA \cite{farhi2014} prepares a variational state by alternating evolution
under the problem Hamiltonian $H_p = H_\text{MVC}$ and a driver (mixing)
Hamiltonian $H_d$,
\begin{equation}
    \label{eq:qaoa_state}
    \ket{\psi_p(\boldsymbol{\gamma},\boldsymbol{\beta})} = \prod_{k=1}^{p} e^{-i \beta_k H_d} e^{-i \gamma_k H_p} \ket{\psi_0},
\end{equation}
where $\ket{\psi_0}$ is the ground state of $H_d$ and the $2p$ angles
$(\boldsymbol{\gamma},\boldsymbol{\beta})$ are either tuned by a classical
outer loop to minimise $\langle H_p \rangle = \bra{\psi_p(\boldsymbol{\gamma},\boldsymbol{\beta})} H_p \ket{\psi_p(\boldsymbol{\gamma},\boldsymbol{\beta})}$ or fixed by a
parameter-free schedule. Measuring the final state in the computational basis
returns candidate vertex covers. In the standard formulation
$H_d = -\sum_{i \in V} X_i$ is a transverse field, whose ground state is the
uniform superposition $\ket{+}^{\otimes \abs{V}}$, so that the circuit explores
the entire $2^{\abs{V}}$-dimensional Hilbert space.
 
All alignment results reported are obtained at depth $p = 3$ with a
linear-ramp schedule~\cite{barrera2025}, both for direct QAOA and for the QAOA
sub-solver called inside the branch-and-bound decomposition, detailed in
Appendix~\ref{app:class}. The schedule fixes the angles rather than optimising
them: the driver (mixer) angle decreases linearly as $\beta_k = 1 - k/p$ while
the problem (cost) angle increases linearly as $\gamma_k = (k+1)/p$ for
$k = 0, \dots, p-1$, an adiabatic-inspired ramp that begins mixer-dominated and
ends cost-dominated. At $p = 3$ this gives
$\beta = (1, \tfrac{2}{3}, \tfrac{1}{3})$ and
$\gamma = (\tfrac{1}{3}, \tfrac{2}{3}, 1)$, and the fully constrained mixer uses
the analogous two-part ramp with an additional initial-Hamiltonian term. Dispensing with the outer loop means that the
differences we report between variants reflect the mixer construction rather
than the behaviour of a classical optimiser, and that the reported runtimes
carry no angle-search cost. The scaling study of Sec.~\ref{sec:results-p1} is a
separate experiment conducted at $p = 1$, where $\expval{H_p}$ is available in
closed form (Appendix~\ref{app:p1}). There the angles are selected as the best
of a $7 \times 7$ grid over $\beta, \gamma \in [0, 2\pi]$ together with the
linear-ramp point. All circuits are simulated with \texttt{qiskit-aer}~\cite{qiskit2024} without
a noise model, drawing $2048$ shots. Ground-state probability is estimated from
these same samples rather than from exact amplitudes, so values below roughly
$5\times10^{-4}$ are not resolved.\\

\paragraph{Constraint handling: penalties versus mixers.}

The vertex-cover constraint can be imposed in two qualitatively different ways.
The penalty route leaves the mixer unconstrained and folds feasibility into the
cost, as in the second term of Eq.~\eqref{eq:h_mvc}: the search visits
infeasible strings, and $\lambda$ controls how strongly they are suppressed.
The mixer route instead restricts $H_d$ so that it is block-diagonal with
respect to the feasible subspace, and initialises the register inside that
subspace, so that the evolution never produces an infeasible string
\cite{hadfield2019,wangrubin2020,baertschi2020}. The two routes trade against
each other directly: penalties are cheap in qubits and depth but return
infeasible samples, whereas structural enforcement guarantees feasibility at
the cost of a more elaborate, deeper mixer. Characterising that trade-off for
this problem is the purpose of the variants below.\\

\paragraph{Clique-cover mixers.}
\label{sec:cliquecover}

Between the two extremes sits a family of mixers built from classical structure
in $\bar{H}$. Given a clique of size $d$, any vertex cover must contain at
least $d-1$ of its nodes. Any disjoint clique cover therefore provides a lower
bound on the minimum vertex cover, since at least all but one node of each
clique must belong to every valid cover. Given a clique cover, we adapt QAOA so
that it only searches over strings consistent with this bound. For each clique
the valid strings consist of the all-\mbox{`1'} state and the one-cold states,
i.e.\ states with a single \mbox{`0'} and \mbox{`1'}s everywhere else, giving
$d+1$ valid states for a clique of size $d$.

\begin{table}[h]
    \centering
    \begin{tabular}{c|c}
         Clique configuration  & Node label \\
         \hline
         011 & 0\\
         101 & 1\\
         110 & 2\\
         111 & 3\\
    \end{tabular}
    \caption{Example valid configurations for a clique of size three. For the one-cold states (left-column) the position of the \mbox{`0'} corresponds to the value associated with the node label, with the first position corresponding to 0. The all-\mbox{`1'} state corresponds to the node label being equal to the size of the clique. }
    \label{tab:clique_labels}
\end{table}

We consider driver Hamiltonians that drive each clique separately and are
block-diagonal, so that they do not couple valid states to invalid ones. Each
choice of driver corresponds to the adjacency matrix of a graph whose nodes are
the valid clique states. Three such graphs are considered here: a \emph{ring},
a \emph{star} graph, and an \emph{all-to-all} graph. Interpreting the one-cold states associated with a clique of size $d$ as nodes labelled by integers between 0 and $d-1$ (a small example is shown in Table~\ref{tab:clique_labels}) and the all-\mbox{`1'} state as the node labelled by $d$, then the graphs represent:
\begin{itemize}
    \item \emph{Ring}: Coupling consecutively labelled nodes, with the addition of coupling node 0 to node $d$.
    \item \emph{Star}: Coupling all nodes to the central node, which is taken to be $d$. This graph is equivalent to the standard Boolean-cube
    connectivity between states induced by transverse-field driving, with the transitions between valid and violating states removed.
    \item \emph{All-to-all}: Coupling valid states to all other valid states. 
\end{itemize}
Throughout, the driver unitaries are implemented exactly
rather than by Trotterisation, so that the results below carry no Trotter error. Appendix~\ref{app:circuits} provides the circuit implementations. 

Because a clique cover constrains only the edges internal to its cliques, the
edges of $\bar{H}$ that run between cliques are still handled by the penalty
term of Eq.~\eqref{eq:h_mvc}. The clique-cover variants are therefore
penalty-based formulations with a partially constrained mixer, intermediate
between the two extremes of the previous subsection. A combination of penalties and constrained mixing has previously been explored for Parity-QAOA \cite{ender2022}. 

Two constructions are used to obtain the clique cover itself. A maximal
matching is an example of a clique cover and can be found in polynomial
time \cite{galil1986}, and for triangle-free graphs it is optimal. We also consider a greedy
algorithm that builds cliques starting from high-degree nodes.\\

\paragraph{The seven variants.}
\label{sec:variants}

\begin{table*}[t]
    \caption{\textbf{The seven QAOA variants.} All variants solve the same problem
    Hamiltonian, Eq.~\eqref{eq:h_mvc}, and differ only in how the vertex-cover
    constraint is enforced. The driver graph is the graph whose adjacency
    matrix defines $H_d$. For the clique-cover variants its nodes are the
    $d+1$ valid states of a clique of size $d$.}
    \label{tab:variants}
    \begin{ruledtabular}
\begin{tabular}{lllll}
        Variant & Constraints & Clique cover & Driver graph & Initial state\\
        \colrule
        \emph{Basic}             & penalty           & ---              & transverse field           & $\ket{+}^{\otimes\abs{V}}$\\
        \emph{Matching Ring}     & penalty $+$ mixer & maximal matching & complete, three states     & uniform, three states\\
        \emph{Matching Star}     & penalty $+$ mixer & maximal matching & star                       & star ground state\\
        \emph{Greedy Ring}       & penalty $+$ mixer & greedy           & ring                       & $W$ state\\
        \emph{Greedy All-to-All} & penalty $+$ mixer & greedy           & all-to-all                 & $W$ state\\
        \emph{Greedy Star}       & penalty $+$ mixer & greedy           & star                       & $W$ state, $H$ on qubit 1\\
        \emph{Constrained}       & mixer             & ---              & vertex covers of $\bar{H}$ & $\ket{1\ldots1}$\\
    \end{tabular}
    \end{ruledtabular}
\end{table*}

Combining the two extremes above with the clique-cover mixers gives the seven variants listed in
Table~\ref{tab:variants}. The \emph{Basic} variant is the penalty-only
baseline: an unconstrained transverse-field mixer acting on the full Hilbert
space \cite{farhi2014}. The \emph{Constrained} variant is the opposite extreme, a mixer
restricted to the valid vertex covers of $\bar{H}$ so that every sample is
feasible by construction \cite{hadfield2019}. The remaining five combine a clique cover
(maximal matching or greedy) with one of the three driver graphs.

The table has five rather than six clique-cover entries because the matching
cover admits no separate all-to-all variant. A maximal matching produces
cliques of size at most two, so each clique register carries at most $d+1 = 3$
valid states, and a ring on three nodes is already the complete graph: the ring
and all-to-all drivers coincide there and define the same unitary. The star on
three nodes is a path rather than a triangle, so \emph{Matching Star} remains
distinct. For the greedy cover, which produces cliques of size $d > 2$, all
three driver graphs differ and all three variants are retained.\\

\paragraph{Branch-and-bound decomposition.}
\label{sec:bnb}
 
Direct QAOA is bounded by the number of available qubits, which for
Eq.~\eqref{eq:h_mvc} is one per candidate-pair node and thus grows as
$\abs{V_1}\cdot\abs{V_2}$ before filtering. In practice this confines direct
solution to instances of at most $33$ qubits. To reach larger instances
we embed each variant in a branch-and-bound decomposition: classical
kernelisation reduces the instance, branching splits it into subproblems of at
most $20$ nodes, and each subproblem is solved by QAOA while larger branches
are resolved classically. Because the bounds used for pruning are valid bounds
on the vertex-cover objective, exactness is preserved throughout.
Appendix~\ref{app:class} provides the details.

\subsection{Benchmarking datasets}
\label{sec:datasets}

\paragraph{Synthetic PPI network pairs.}

We generated synthetic PPI network pairs using NAPAbench2~\cite{woo2020}, which produces evolutionarily related networks from a common ancestor along a hypothetical phylogenetic tree, with known ground-truth node correspondences. Networks were extended using the STICKY growth model~\cite{przulj2006} as implemented in NAPAbench2. We considered two ancestor sizes, corresponding to two evolutionary scenarios, and generated five independent pairs per parameter setting. Node weights were taken from the sequence-similarity matrices supplied by NAPAbench2 alongside each network pair. The resulting pairs span a range of sizes and densities, from the smallest instances up to the largest ones that remain tractable for the QAOA variant simulations, and are summarised in Table~\ref{tab:napabench}. Within each pair the two networks differ in size and density, reflecting the asymmetry typical of cross-species PPI comparisons.\\

\paragraph{Real-world PPI network pairs.}

We complemented the synthetic benchmarks with real PPI network pairs derived
from the human and mouse interactomes in STRING \cite{szklarczyk2023}, retaining interactions with combined score $\geq$ 700 (high confidence). Cross-species pairs were
constructed around shared KEGG~\cite{kanehisa2000} pathways: for each pathway
present in both organisms, we induced the subgraph of the corresponding pathway proteins in each species' interactome. Modular product graph node weights were computed with
BLAST~\cite{altschul1990,camacho2009}: hits were generated with NCBI
BLAST+ \texttt{blastp} 2.16.0 at an $E$-value threshold of $10^{-5}$ against
the STRING v12 protein sequences~\cite{szklarczyk2023}, and the weights are the resulting bit-scores.
The pathway pairs used are listed in Table~\ref{tab:keggpathways} and shown in Fig.~\ref{fig:kegg_dataset}.\\

\paragraph{Disease-associated genes.}

To ground the alignments in disease-relevant biology, we annotated the pathway networks
with genes associated with Alzheimer's and Parkinson's disease. 
Disease genes were obtained from the Open Targets Platform~\cite{ochoa2023, buniello2025}, retaining for
each disease all targets with a genetic-association datatype score $\geq 0.1$ (release 26.06), following the threshold used by Federico et al.~\cite{federico2022}.

\subsection{Benchmarking measures}
\label{sec:measures}

We assess alignment quality along two axes, topological and biological
conservation. Both are reported alongside node coverage, since either score can
be inflated by aligning fewer nodes.\\

\paragraph{Topological conservation.}

As the primary structural measure we use the Symmetric Substructure Score
($\Sthree$) \cite{saraph2014}. An edge of the subgraph induced on the mapped
nodes of the first network is conserved when its image is an edge of the
second. $\Sthree$ is the Jaccard index of the two induced edge sets,
\begin{equation}
    S^{3} = \frac{|E_c|}{|E_1^{\text{ind}}| + |E_2^{\text{ind}}| - |E_c|},
\end{equation}
where $E_c$ is the set of conserved edges and $E_i^{\text{ind}}$ the edges
induced on the mapped nodes $V_i'$ in the network $i$. Unlike the one-sided
Edge Correctness \cite{kuchaiev2010} and Induced Conserved Structure
\cite{patro2012}, of which it is the symmetric combination, $\Sthree$
penalises both sparse-to-dense and dense-to-sparse mismapping.\\

\paragraph{Biological conservation.}

For synthetic benchmarks, where the true orthology
mapping is known, we report node correctness as node-level precision: the
fraction of evaluable mapped nodes assigned to their true counterpart~\cite{woo2020}.

For real networks, where no ground-truth mapping exists, we use Gene Ontology
(GO)~\cite{ashburner2000,go2026} semantic similarity over the biological-process
namespace. Annotations are taken from the UniProt-GOA human and mouse
releases~\cite{huntley2015} and stored per pathway as gene-to-term lists,
retaining all evidence codes. The ontology is read from the \texttt{go-basic}
release, restricted to biological-process terms connected by \texttt{is\_a} and
\texttt{part\_of} edges. For a term $t$, the information content is
$\mathrm{IC}(t) = -\log p(t)$~\cite{resnik1995}, with $p(t)$ the frequency of $t$ and its
descendants in the annotation corpus. Two terms $t_1, t_2$ are scored with the
Lin measure~\cite{lin1998},
\begin{equation}
  \mathrm{sim}(t_1, t_2) =
  \frac{2\,\mathrm{IC}(t_\mathrm{MICA})}{\mathrm{IC}(t_1) + \mathrm{IC}(t_2)},
\end{equation}
where $t_\mathrm{MICA}$ is the common ancestor of $t_1$ and $t_2$ with the
highest information content, and the score is zero if the denominator vanishes.
Term-level scores are aggregated per gene pair by the Best-Match-Average
scheme~\cite{schlicker2006}: for each term of one protein the maximum
similarity to any term of its partner is taken, and the two directional means
are averaged. The pathway-level score is the mean over aligned pairs. Pairs in
which either protein lacks a biological-process annotation are excluded from
this mean.\\

\paragraph{Node coverage.}

Because every score above is interpretable only relative to how much of the
network is mapped, we report node coverage
\begin{equation}
    \mathrm{NCV} = \frac{|V_1'|}{|V_1|},
\end{equation}
the fraction of the smaller network that the alignment matches.

\subsection{Classical baselines}
We benchmarked against seven classical global network aligners: L-GRAAL~\cite{maloddognin2015},
HubAlign~\cite{hashemifar2014}, IsoRank~\cite{singh2008}, MAGNA++~\cite{vijayan2015},
Natalie2~\cite{elkebir2015}, SANA~\cite{mamano2017} and AntNetAlign~\cite{rodriguez2023}. Every aligner was
run once per instance on both benchmark suites (the synthetic NAPAbench2
networks and the human--mouse KEGG pathway pairs), using its published default
parameters. Sequence similarity was supplied to each tool in its expected form:
raw BLAST bit scores for L-GRAAL, HubAlign and SANA, min--max-normalised scores
for MAGNA++, IsoRank and AntNetAlign, and E-values for Natalie2.

\section{Results}
\label{sec:results}

\subsection{QAOA variants on small instances}
\label{sec:results-small}

\begin{figure}[htb]
    \centering
    \includegraphics[width=\columnwidth]{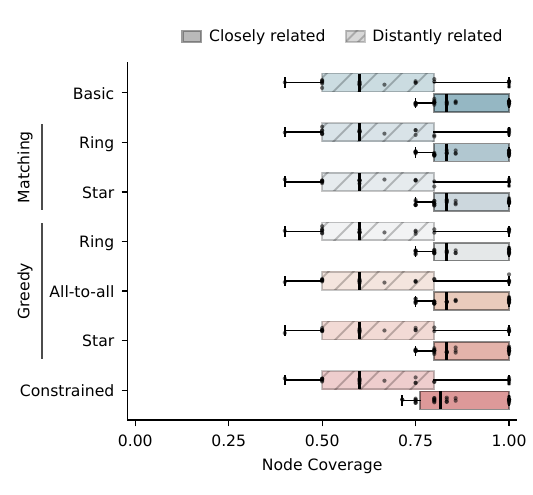}
    \caption{\textbf{Node coverage of the QAOA variants on the small
    NAPAbench2 instances.} Solid boxes are closely related pairs, hatched
    boxes distantly related ones. Colours run from blue to red as feasibility moves from the penalty term (\emph{Basic}) into the mixer (\emph{Constrained}).}
    \label{fig:coverage}
\end{figure}

We first benchmark the seven QAOA variants on the small synthetic NAPAbench2 pairs of maximal size of $33$ modular product size after filtering, which are small enough to be solvable in a single QAOA run, without classical decomposition. \\

\paragraph{Alignment quality.}
Across all seven variants the mapped core is conserved exactly. Node correctness is $1.0$ for all variants, and topological conservation is $1.0$ for both closely and distantly related pairs. The latter follows from the induced MCS objective: every induced edge of the mapped core is conserved by construction.
Node coverage does not separate the variants either (Fig.~\ref{fig:coverage}): all seven yield near-identical distributions in both regimes. The constraint-handling strategy therefore has no measurable effect on alignment quality at these instance sizes. What coverage does depend on is how divergent the two networks are: medians fall from approximately $0.82$ for closely related pairs to approximately $0.60$ for distantly related ones, as fewer nodes can be matched when the networks share less structure.\\

\paragraph{Hardware cost and sampler behaviour.}
\begin{figure*}[htb]
    \centering
    \includegraphics[width=\linewidth]{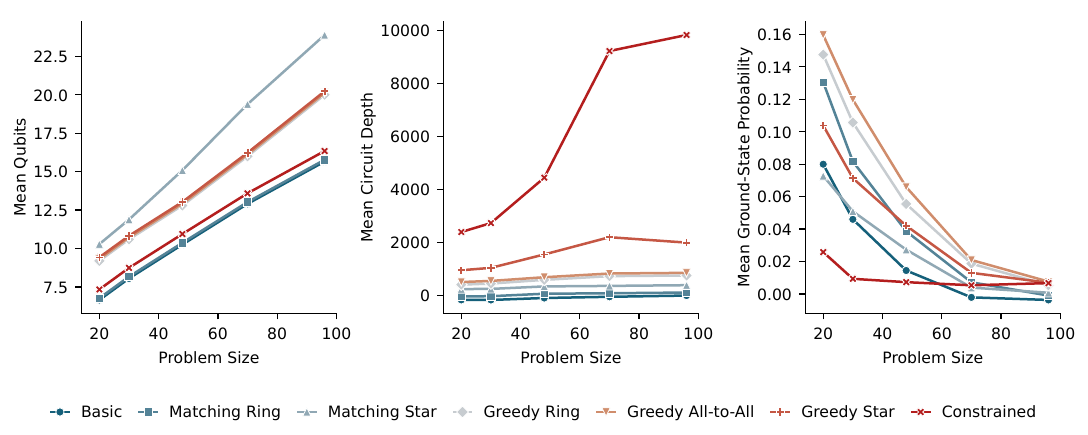}
    \caption{\textbf{Hardware cost and sampler behaviour of the QAOA variants on closely related synthetic pairs}.
    Qubit count, circuit depth
    (decomposed to a \{u,\,cx\} basis), and ground-state probability, the
    fraction of samples that hit the exact optimum with increasing problem size. Colours run from blue to red as feasibility moves from the penalty term (\emph{Basic}) into the mixer (\emph{Constrained}).}
    \label{fig:scaling}
\end{figure*}

Both the hardware cost of the variants and the quality of their samplers change systematically with problem size (Fig.~\ref{fig:scaling}). Qubit count grows sub-linearly in the nominal size, where without kernelisation it would be linear, and the remaining kernel splits the variants into three groups. \emph{Basic}, \emph{Matching Ring} and \emph{Constrained} sit at
the one-qubit-per-variable baseline. The greedy clique-cover variants lie above them, with one auxiliary qubit per clique register. \emph{Matching Star} is the
most expensive, because a maximal matching produces many
small cliques, each needing an auxiliary qubit (Appendix~\ref{app:circuits}). Circuit
depth separates them more, driven by the fully constrained variant:
enforcing feasibility directly in the mixer rather than through penalty terms
lifts its depth to a mean of roughly $9{,}500$ at the largest instance, five times
that of \emph{Greedy Star} (roughly $1{,}800$) and more than an order of magnitude
above the rest, which stay below $800$. These circuits are noiseless, but depth sets the order in which the variants would accumulate error on hardware. The feasibility guarantee of the constrained mixer holds only for the ideal circuit, so whether the structured mixers keep their advantage under noise is open.
Ground-state probability, the fraction of samples that hit the exact optimum,
stays low throughout and decays towards what $2048$ shots resolve at the largest
instances: on closely related pairs the penalty-based variants reach
means of roughly $0.08$--$0.16$ at the smallest size (highest for the greedy ring
and all-to-all drivers) and fall below $0.02$ by problem size $70$, so no variant
reliably samples the exact optimal cover under QAOA. The fully
constrained variant is in fact the weakest on all but the largest closely related pairs, so restricting the mixer to the feasible subspace does not, at
this depth, concentrate the sampler on the optimum.

\subsection{Branch-and-bound and QAOA variants at scale}
\label{sec:results-bnb}

\begin{figure*}[htb]
    \centering
    \includegraphics[width=\linewidth]{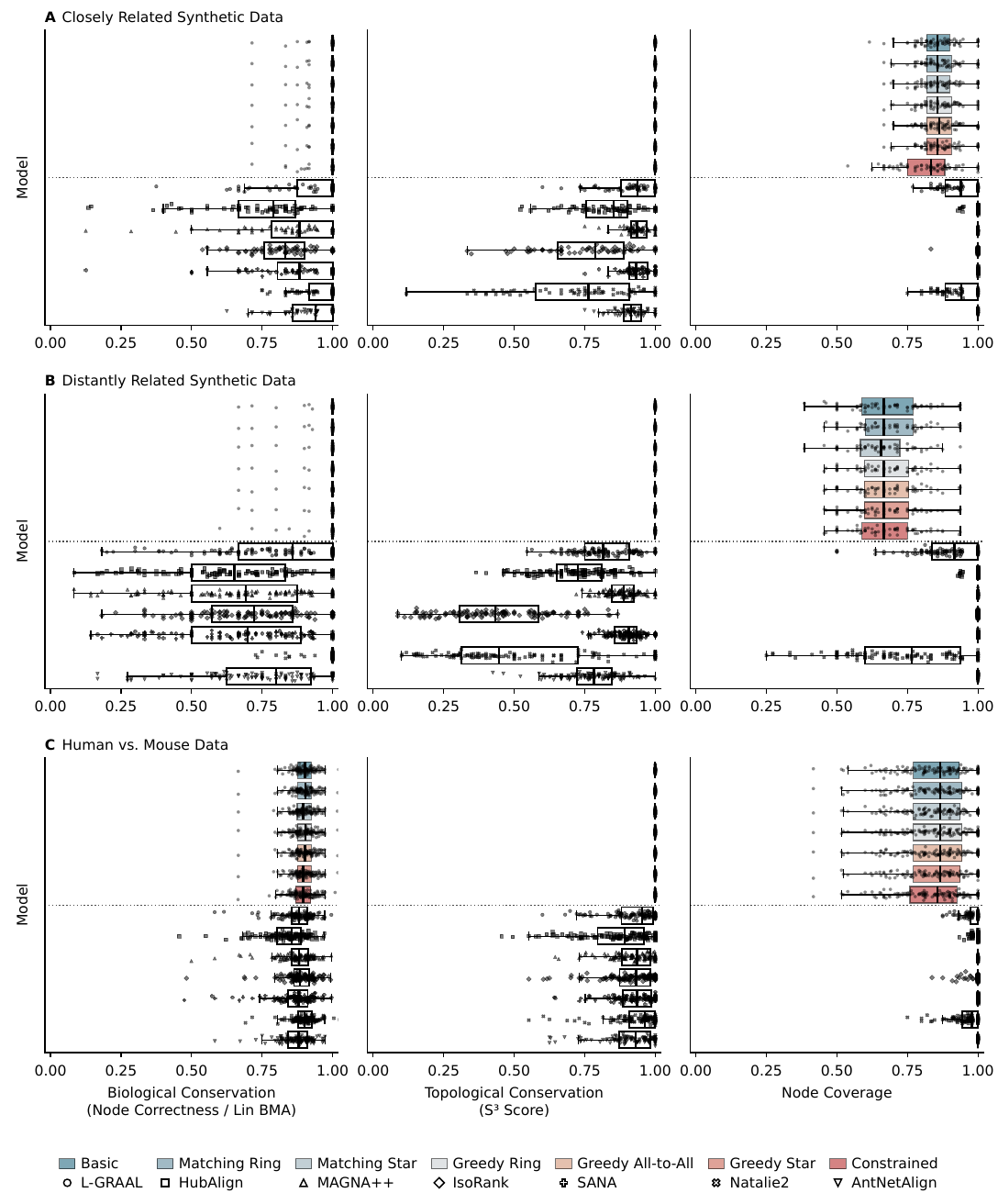}
    \caption{\textbf{Alignment quality of the branch-and-bound decomposition combined with the QAOA
    variants compared against the classical aligners.} Each panel shows the
    distribution of alignment quality over network pairs for one metric:
    biological conservation (node correctness for synthetic data, Lin
    best-match-average GO semantic similarity for real-world data), topological
    conservation ($\Sthree$ score), and node coverage. Rows correspond to (A)
    closely related and (B) distantly related synthetic NAPAbench2 pairs, and
    (C) real human--mouse KEGG pathway pairs. Colours run from blue to red as feasibility moves from the penalty term (\emph{Basic}) into the mixer (\emph{Constrained}).}
    \label{fig:bnb-classical}
\end{figure*}

To move beyond the instance sizes reachable by direct QAOA, we embed each
variant in the branch-and-bound decomposition, which
partitions an instance into qubit-feasible subproblems while preserving
exactness. This is a substantial increase over direct QAOA, though still well below whole-interactome scale. We benchmark the resulting branch-and-bound $+$ QAOA pipeline against established
classical aligners on both synthetic and real network pairs
(Fig.~\ref{fig:bnb-classical}).\\

\paragraph{Synthetic benchmarks.}
On closely related NAPAbench2 pairs all seven variants attain node correctness
of $1.0$ for nearly every pair, with $\Sthree$ at $1.0$ by construction of the
induced objective (Fig.~\ref{fig:bnb-classical}A). The classical aligners score lower on both measures and
scatter considerably more, several distributions extending far into the low
range. What the quantum variants give up is node coverage, at a median of
roughly $0.85$ against near-complete coverage for most classical methods.
Natalie2 and L-GRAAL are the exceptions, since both may leave nodes unmatched
rather than forcing a full assignment, and their coverage is only slightly higher than ours.

The separation grows on the distantly related pairs (Fig.~\ref{fig:bnb-classical}B). As the networks diverge
the classical aligners degrade on both axes, node correctness shifting lower
and $\Sthree$ spreading well below ceiling, while the branch-and-bound $+$ QAOA variants
remain at ceiling throughout. Coverage declines to about $0.65$ for our
variants and drops for Natalie2 and L-GRAAL alongside them, since fewer nodes
admit an exact match once the two networks share less structure. The classical
aligners were designed for interactomes orders of magnitude larger than these
instances.\\

\paragraph{Real-world benchmarks.}
On the human--mouse KEGG pathway pairs $\Sthree$ is again at ceiling across all
variants, and GO semantic similarity is comparable between the two groups, with
medians clustered around $0.85$--$0.9$ (Fig.~\ref{fig:bnb-classical}C). Coverage again favours the classical methods. Since Natalie2 and L-GRAAL produce almost complete alignments in this case, the lower coverage appears to result from our formulation rather than from partial mapping itself.

The cause is the induced rule, which admits a pairing only where the
interaction is present in both species or absent in both. Human and mouse
subgraphs are not equally dense and not equally well studied \cite{schaefer2015}. An
edge present on one side and absent on the other vetoes a pairing irrespective
of which side it sits on, so the sparser network bounds how much of the denser
one can be mapped.

Two relaxations address this, and both gain coverage at the expense of exact
conservation. Weakening the rule to a one-way implication
(Appendix~\ref{app:mono}) raises coverage close to the ceiling, while $\Sthree$ decreases (Fig.~\ref{fig:appendix-mono-bnb}).
Lowering the penalty weight below $\max_i w_i + 1$ (Appendix~\ref{app:relaxed}, Fig.~\ref{fig:penalty-sweep})
allows a bounded number of uncovered edges, which costs more $\Sthree$ for less
coverage and adds a tail of poorly conserved pairs
(Fig.~\ref{fig:appendix-threeway}). Here feasibility no longer follows from
optimality there, the pruning bounds break.

\subsection{Use case: Human--mouse alignment of pathways associated with
Alzheimer's and Parkinson's disease}
\label{sec:usecase}

\begin{figure*}[htb]
    \centering
    \includegraphics[width=\linewidth]{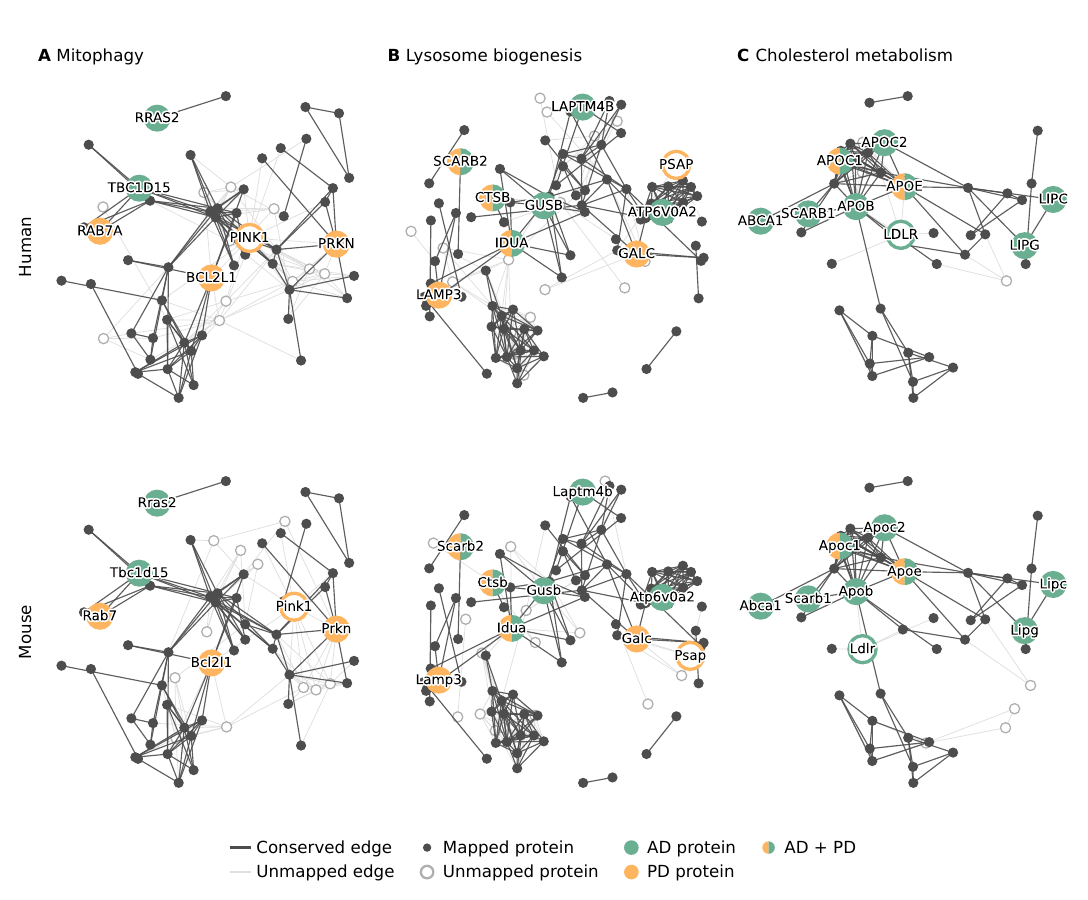}
    \caption{\textbf{Cross-species alignment of disease-protein neighbourhoods in three
    neurodegeneration-associated pathways.} Human and mouse
    STRING networks for (A) Mitophagy (KEGG hsa04137), (B) Lysosome biogenesis (hsa04142) and
    (C) Cholesterol metabolism (hsa04979). Proteins whose genes are associated with
    Alzheimer's disease (AD) in Open Targets are shown in green, those associated with
    Parkinson's disease (PD) in yellow. Proteins aligned to a
    counterpart by the \emph{Matching Ring} QAOA variant are drawn at matching
    positions in the two species' layouts.}
    \label{fig:usecase}
\end{figure*}

We examine three of the aligned KEGG pathway pairs in more detail, each with an
established role in neurodegenerative processes implicated in Alzheimer's (AD)
and Parkinson's disease (PD).
Mitophagy (hsa04137) mediates the selective clearance of damaged mitochondria. Nigral dopaminergic
neurons sustain an unusually high bioenergetic and oxidative load, and defects in
mitochondrial quality control have long been associated with their selective
vulnerability in PD~\cite{pickrell2015,ge2020}. Lysosome biogenesis (hsa04142) produces the
terminal degradative compartment for autophagy and endocytosis, and lysosomal dysfunction
can therefore impair the clearance of damaged organelles and aggregation-prone substrates
that accumulate in both diseases~\cite{robak2017,nixon2024}. Cholesterol metabolism
(hsa04979) regulates lipid homeostasis and transport in the brain, where altered lipid
handling is strongly implicated in AD~\cite{yamazaki2019,bellenguez2022}. These pathways
are functionally coupled: mitochondria targeted for mitophagy are ultimately degraded in
lysosomes, and altered cholesterol homeostasis acts on the same endolysosomal compartment
and can itself interfere with mitochondrial clearance~\cite{deus2020,nixon2024}. Although
the underlying genes are conserved at the orthologue level, their phenotypic consequences
can differ between human and mouse~\cite{kitada2009,sullivan1997}.

All three pairs are solved through the branch-and-bound decomposition, and five of the seven implemented QAOA variants recover the identical alignment on each. The others differ only in paralogue assignment: \emph{Basic} cross-assigns one paralogue pair each in mitophagy and the lysosome pair, and \emph{Constrained} aligns two proteins fewer in cholesterol metabolism and one fewer in the lysosome pair. Fig.~\ref{fig:usecase} shows the \emph{Matching Ring} solution, with the disease proteins highlighted.

Across the three pathways, $21$ of the $24$ disease proteins ($88\%$) present in
the human networks are recovered by the induced alignment, compared with $151$ of $181$
($83\%$) of the remaining proteins. Viewed from the opposite perspective,
disease proteins make up a larger fraction of the mapped proteins ($21/172 = 12.2\%$)
than of the unmapped proteins ($3/33 = 9.1\%$). The number of disease
proteins is small, but the direction is
consistent: the proteins of primary biological interest are recovered at
least as well as, and, if anything, better than the remaining network proteins,
even though the strict induced formulation yields lower overall node coverage than the
classical aligners. This is noteworthy because the
induced rule maps a node only when its interactions are matched on both sides, making
recovery increasingly demanding for highly connected nodes. Several of these proteins are
among the better-connected nodes of their pathway networks, consistent with the tendency
of disease proteins to occupy connected neighbourhoods of the
interactome~\cite{menche2015}.

Among the recovered proteins are several with particularly strong genetic evidence for
disease in the respective pathways. \textit{PRKN}, which is mutated in
autosomal-recessive early-onset PD~\cite{kitada1998}, is recovered in the mitophagy panel, as is
\textit{APOE}, a major common genetic risk factor for late-onset AD~\cite{yamazaki2019}, in the cholesterol
panel. \textit{APOE} additionally illustrates that the disease annotations overlap at the
gene level: it is associated with both AD and PD, as are \textit{SCARB2}, \textit{CTSB},
and \textit{IDUA} in the lysosome panel. The three disease proteins not recovered are
\textit{PINK1}, \textit{PSAP}, and \textit{LDLR}. \textit{PINK1} is the most notable of
these, since it acts upstream of \textit{PRKN} in the same pathway~\cite{pickrell2015}. Its human interaction
neighbourhood is not sufficiently matched in the mouse network to satisfy the
induced-alignment constraint.

The alignment moreover places the recovered proteins on their corresponding mouse orthologues: $20$ of the $21$ map to identically named mouse genes, while the remaining gene maps to its orthologue under a different symbol (\textit{RAB7A} $\rightarrow$ \textit{Rab7}). 
The conservation extends to
their surroundings: of the $76$ distinct first-order neighbours of the mapped disease
proteins, $67$ ($88\%$) are themselves mapped ($79$--$94\%$ per pathway), and $63$ of the
$71$ orthology-resolvable neighbours ($89\%$) are aligned to the correct orthologue.
Because the induced rule preserves adjacency, these neighbours retain their interaction
with the disease protein on the mouse side. They are also functionally coherent, with a Lin
best-match-average GO biological-process similarity of $0.86$--$1.00$ across the three
pathways. The alignment thus preserves not only the disease proteins but the conserved
subnetwork around them.

\subsection{Depth-one QAOA variants at scale}
\label{sec:results-p1}
 
\begin{figure*}[htb]
    \centering
    \includegraphics[width=\linewidth]{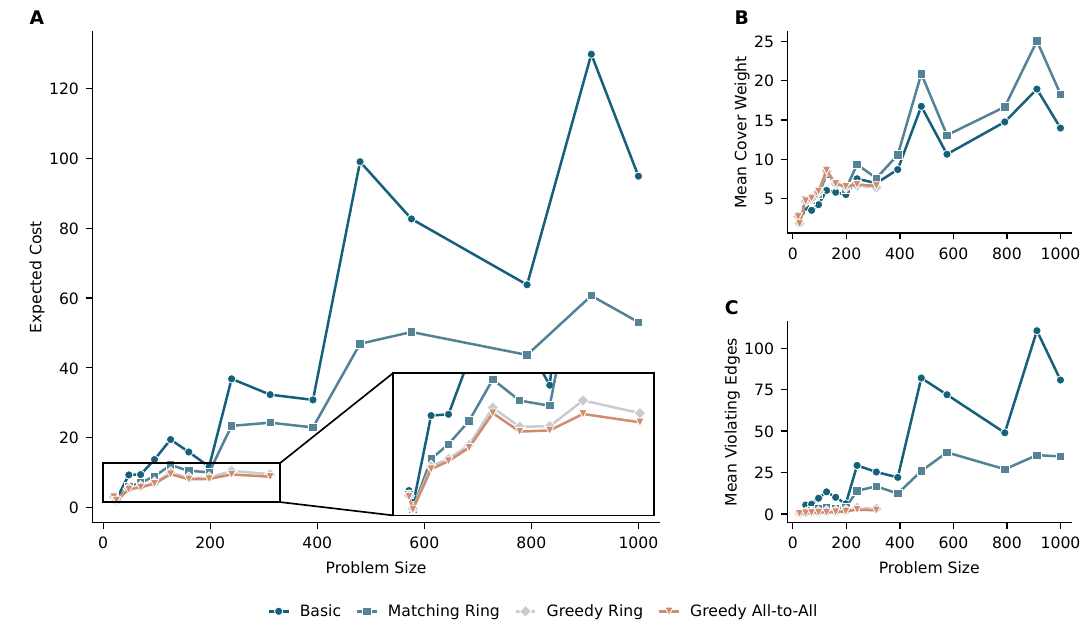}
    \caption{\textbf{Depth-one QAOA cost decomposition on closely related synthetic
    networks.} For each of the four variants whose depth-one expectation is
    available in closed form, the (A) exact expected cost
    $\expval{H_p} = \expval{C} + w_\text{max}\expval{P}$ is split into (B) the average vertex-cover weight $\expval{C}$ and (C) the average number of
    uncovered edges $\expval{P}$, against the modular-product size
    $\abs{V_1}\cdot\abs{V_2}$. The maximum node weight is denoted by $w_\text{max}$.
    Colours run from blue to red as feasibility moves from the penalty term (\emph{Basic}) into the mixer (\emph{Greedy}).}
    \label{fig:p1}
\end{figure*}
 
The comparison in Sec.~\ref{sec:results-small} is bounded by simulation cost to
instances of at most $33$ qubits. For the four variants whose mixer is
circulant the depth-one expectation of the cost Hamiltonian is instead available
in closed form (Appendix~\ref{app:p1}), which removes that ceiling. Its
evaluation cost grows with the size of the largest clique in the
cover, so the reach follows clique size: \emph{Basic} and \emph{Matching Ring} can be
characterised up to modular product sizes of $1000$ and the greedy variants stop at $380$ (Fig.~\ref{fig:p1}).

The reach comes at a price. A single layer yields the expected cost
$\expval{H_p}$ but no sampled bitstrings, so it bounds the achievable alignment
rather than reporting one. In return the closed form is exact, free of sampling
noise, and shows $\expval{H_p}$ alongside the two quantities that drive it: the
weight of the vertex cover a variant selects and the number of edges it leaves
uncovered.

The greedy variants work best, keeping the expected cost below $15$ across the sizes they reach, whereas it rises to $60$ for \emph{Matching Ring} and to $130$ for \emph{Basic} (Fig.~\ref{fig:p1}A).

The first of the two drivers is the cover weight (Fig.~\ref{fig:p1}B). A lighter
cover is better, since it leaves more of the sequence-similarity weight, and typically more nodes, in the
alignment. \emph{Basic} selects the lightest covers over the small-size range,
with the clique-structured mixers marginally heavier. \emph{Matching Ring} stays
just above \emph{Basic} almost throughout, and the \emph{Greedy Ring} and
\emph{Greedy All-to-All} curves cross the \emph{Basic} curve.

At depth one a light cover can equally be an infeasible one, and \emph{Basic}
reaches its light covers by leaving edges uncovered (Fig.~\ref{fig:p1}C).
\emph{Basic}, whose mixer imposes no cover-feasibility structure, leaves a large
and growing expected number of edges uncovered, peaking near $110$. The
structured mixers suppress this, with the greedy variants staying close to zero
across the range they reach and \emph{Matching Ring} in between, ending near
$35$. The
trajectory of \emph{Basic} is markedly non-monotonic, reflecting
instance-to-instance variation among the network families at a given
product-graph size.

Overall, the greedy variants reach the lowest expected cost but only over the
smaller instances their closed form can evaluate, whereas \emph{Basic} spans the
full range at the highest cost, driven by its many uncovered edges. \emph{Matching
Ring} strikes the best balance, holding a cost well below \emph{Basic} across the
full range that \emph{Basic} reaches.

\clearpage
\section{Discussion and conclusion}
\label{sec:conclusion}

We reduce protein--protein interaction network alignment to a weighted minimum
vertex cover on the complement of the modular product graph and solve it with
seven QAOA variants embedded in a branch-and-bound decomposition. The variants
differ only in how the vertex-cover constraint is enforced.

Alignment quality comes out the same for all seven variants. The exact reduction holds
both conservation measures at ceiling across the range we tested, including the
distantly related pairs where less structure is shared. What we give up is
coverage, and on the human--mouse pathways the induced rule is what binds, since
an interaction recorded in one species and missing in the other blocks the
pairing outright. Both relaxations we tested buy coverage back at the expense of
exact conservation, and they differ in what else they cost: the one-way
implication leaves the branch-and-bound decomposition intact, while a lowered
penalty weight forfeits the pruning bounds.
The smaller mapped subset nevertheless retains what matters biologically, with
the disease-associated proteins of the pathways kept and mapped to their
orthologues.

What separates the variants is their cost. The fully constrained mixer is the
only formulation that guarantees feasibility, and it is also the most expensive
by a wide margin. The clique-cover mixers enforce part
of the constraint and pay for it in auxiliary qubits, most of all \emph{Matching Star},
while \emph{Matching Ring} stays at the one-qubit-per-variable baseline. \emph{Basic} is the
cheapest. At depth one, where the closed form carries the comparison to modular
product graphs an order of magnitude larger, the same ordering appears in the
expected cost: \emph{Basic} leaves a large and growing number of violated edges,
\emph{Matching Ring} fewer, and the greedy covers stay near zero, while the average
cover weight separates them only weakly. The clique-cover variants therefore sit
between the two extremes, enforcing part of the constraint at a fraction of the
depth the constrained mixer requires.

These results have several limitations. The instances we align are far below
interactome scale. On instances of this size, classical exact solvers find the
optimal alignment efficiently, so what we contribute is the encoding and its
resource cost, not direct evidence of a computational speedup. The classical aligners we benchmark
against were built for whole interactomes, so their results are not optimal
here. The constrained variant, the only one whose mixer guarantees feasibility,
requires a circuit depth far beyond the reach of near-term hardware, and even
the cheaper variants inherit the general obstacles of QAOA at scale, with
ground-state probability low throughout and falling below the sampling
resolution at the largest sizes we reach. Our angles are
fixed by a linear-ramp schedule rather than optimised, which isolates the effect
of the mixer construction. The depth-one analysis is bounded in turn by the cost
of evaluating the closed form, which grows exponentially in the size of the
largest clique in the cover. The sizes each variant reaches there reflect that
evaluation cost, not the resources the variant itself would need.

The most promising direction forward is to relax exact edge conservation, though
that needs a different decomposition method, since the soft penalty breaks the
bounds branch-and-bound relies on. Another direction is to use the whole sample
rather than the single best mapping: QAOA returns many low-energy covers, and
the near-optimal ones are alternative alignments that could be merged into a
many-to-many mapping, or used in other ways we have not explored here. The
formulation could also be extended from pairwise to multiple network alignment.

\section*{Data availability}

All primary data are public. Human and mouse interactomes were taken from STRING v12
\cite{szklarczyk2023} and restricted to KEGG pathways \cite{kanehisa2000}. Synthetic network
pairs were generated with NAPAbench2 \cite{woo2020}. Disease--gene associations are from the
Open Targets Platform \cite{ochoa2023}.

\section*{Acknowledgements}
This work was funded by the Investitions- und Förderbank Hamburg (IFBHH, Hamburg Investment and Development Bank) [51172434].

Graphical abstract and Fig.~\ref{fig:pipeline} are created in BioRender. Stahl, M. (2026) \url{https://BioRender.com/sa0gkgx}.

During the preparation of this work, the authors used Claude (Anthropic) to generate and review snippets of text and code, and to assist with proofreading and improving both. All methods and interpretations of the results were conceived and determined by the authors. After use, the authors reviewed and edited all AI-assisted content as necessary and take full responsibility for the content of the published article. AI was not used as, and is not credited as, an author of this work.

\newpage
\begin{widetext}

\appendix
\setcounter{figure}{0}
\setcounter{table}{0}
\renewcommand{\thefigure}{S\arabic{figure}}
\renewcommand{\thetable}{S\arabic{table}}

\section{Kernelisation and Branch-and-bound}
\label{app:class}

 Kernelisation in this context, refers to applying a series of simple rules (or equivalently passes) to remove the easy part of the problem leaving behind the kernel or difficult part of the problem to solve.

 Each pass in the kernelisation takes as an input an instance of a minimum vertex cover and then returns a new instance of minimum vertex cover with potentially a partial solution to the problem. The returned problem instance should be smaller or the same size in terms of the number of nodes as the original problem. By solving the kernelised problem and combining it with the partial solution, the solution to the original problem should be obtained. The kernelisation should also run in polynomial time. 
Kernelisation has been explored previously in the quantum optimisation literature \cite{schuetz2026, pelofske2019, pelofske2019_2, pelofske2020, pelofske2023}.  In this work we implement the following passes:
 \begin{enumerate}
     \item Remove degree 0 nodes.
     \item Remove degree 1 nodes whose weight exceeds or is equal to its neighbour.
     \item Remove corner nodes. A node is a corner node if it forms a clique with all its neighbours and has weight greater than or equal to its neighbours. Any one of its neighbours covers the same or greater number of edges at less cost \cite{fomin2009, akiba2014}.
     \item Add dominating nodes. A node $u$ dominates a node $v$ if the closed neighbourhood of $v$ is a subset of the closed neighbourhood of $u$. A closed neighbourhood being all neighbours of the node as well as the node itself. Therefore $u$ covers all the edges $v$ covers and more. Provided the weight of $u$ is less than or equal to $v$, then $u$ can always be added to the vertex cover \cite{schuetz2026, butenko2002}.
 \end{enumerate}
This does not present an extensive list of kernelisation techniques \cite{fomin2009, iwata2013, akiba2014}.

For the branch-and-bound methods considered in this paper, the lower bound is
calculated using the maximum matching on the graph. An initial upper bound is
found using a greedy solver that adds nodes, scored by the sum of the weights of
their neighbours, in decreasing order until a vertex cover is reached. The
branch-and-bound branches on high-degree nodes first, with ties broken by the
node whose neighbourhood has the greatest weight. Once a branch has reached a
predefined size, it is passed to QAOA to solve the subproblem.

\section{Details of QAOA \texorpdfstring{$p=1$}{p=1} performance}
\label{app:p1}

In this section we provide the details on analysing QAOA $p=1$ applied to
minimum vertex cover. As the circuits are shallow in this regime we can
evaluate expectation values analytically (this has already been done for the basic version of QAOA $p=1$ \cite{ozaeta2022}). To do this we start by rewriting the
problem Hamiltonian in terms of qudits (i.e.\ arbitrary finite-level quantum systems).
We do this by assigning each clique in the clique covering a single qudit. So
for a clique cover $\mathcal{C} = \{ c_1, c_2 \dots c_m\}$ consisting of $m$
cliques, each clique indexed by $i$ of size $d_i$ is treated as a $d_i+1$ level
qudit. The qudit levels are indexed between $0$ and $d_i$. An index $j$ between
$0$ and $d_i-1$ corresponds to omitting the corresponding qubit from the vertex
cover. The index $d_i$ corresponds to all qubits in the clique being in the
vertex cover. Table~\ref{tab:enc_qudits} shows an example for a clique of size
4.

\begin{table}[h]
    \centering
    \begin{tabular}{c|c}
         Qubit state & Qudit State \\
         \hline
         0111 & 0\\
         1011 & 1\\
         1101 & 2\\
         1110 & 3\\
         1111 & 4\\
    \end{tabular}
    \caption{An example encoding from a clique of size four into a 4-level qudit.}
    \label{tab:enc_qudits}
\end{table}

In this new formulation the problem Hamiltonian, Eq.~\eqref{eq:h_mvc}, can be
written as follows:
\begin{equation}
    \label{eq:hp_clique}
    H_p = \sum_{i=1}^m\sum_{j=0}^{d_i} \left(W_i - w_{ij}\right)P_i(j)  + \lambda \sum_{i_1=1}^m\sum_{i_2=1}^m \sum_{j_1=0}^{d_{i_1}-1}\sum_{j_2=0}^{d_{i_2}-1} \delta_{((i_1, j_1),(i_2, j_2))\in E}P_{i_1}(j_1) P_{i_2}(j_2)
\end{equation}
In Eq.~\eqref{eq:hp_clique} the operators $P_{i}(j)$ is the projector associated
with the $i^{\text{th}}$ clique, projecting onto the $j^{th}$ qudit
computational basis-state. The weights are denoted by $w_{ij}$, i.e.\ the cost
of having the $j^\text{th}$ node in the $i^\text{th}$ clique in the solution.
The set of weights is extended such that $w_{i,d_i} = 0$. The sum of weights
for each clique $i$ is given by
 \begin{equation}
     W_i = \sum_{j=0}^{d_i} w_{ij}.
 \end{equation}
Finally, $\delta$ denotes the Kronecker-delta, which takes the values $1$ or
$0$. In this case $\delta_{((i_1, j_1),(i_2, j_2))\in E}$ is one if there is an
edge in the original vertex-cover graph between the $j_1^\text{th}$ node
associated with the $i_1^\text{th}$ clique and the $j_2^\text{th}$ node
associated with the $i_2^\text{th}$ clique.

It remains to rewrite the driver Hamiltonian in this qudit language. For this
we consider only the case where the adjacency matrix corresponds to a circulant
graph (e.g.\ all-to-all or a ring as opposed to a star graph).

Circulant adjacency matrices can be written in terms of the shift operator
defined as
\begin{equation}
    X_i \ket{k} = \ket{k+1 \mod (d_i+1)},
\end{equation}
where $X_i$ acts on a qudit of dimension $d_i+1$. Therefore, the driver
Hamiltonian can be written as
\begin{equation}
    \label{eq:hd_clique}
    H_d = -\sum_{i=1}^{m}\sum_{j=0}^{d_i} \Gamma_{ij} X_{i}^j,
\end{equation}
with $\Gamma_{ij}\geq 0$ (explicit expressions for the ring and all-to-all drivers can be found in Eq.~\eqref{eq:ring_gamma} and Eq.~\eqref{eq:all_gamma} respectively). The ground state is then $\bigotimes_{i=1}^m
\ket{+}_i$, where $\ket{+}_i$ denotes a uniform superposition of computational
basis states for each qudit $i$.

The aim is to calculate:
\begin{equation}
    \langle H_p \rangle = \bra{+} e^{i\gamma H_p} e^{i\beta H_d} H_p e^{-i\beta H_d} e^{-i\gamma H_p} \ket{+},
\end{equation}
to do this we calculate
\begin{equation}
    \langle Z^r_k \rangle = \bra{+} e^{i\gamma H_p} e^{i\beta H_d} Z^r e^{-i\beta H_d} e^{-i\gamma H_p} \ket{+},
\end{equation}
and
\begin{equation}
    \langle Z^r_kZ^s_l \rangle = \bra{+} e^{i\gamma H_p} e^{i\beta H_d} Z^rZ^s e^{-i\beta H_d} e^{-i\gamma H_p} \ket{+},
\end{equation}
where $Z_i$ is the clock-operator of a $d_i+1$ level qudit defined as
\begin{align}
    Z_i \ket{k} = \omega_{d_i+1}^k \ket{k},
\end{align}
where $\omega_{d_i+1} = e^{2 \pi i /(d_i+1)}$ is the $(d_i+1)^\text{th}$ root
of unity.

Calculating the expectation values involves the following steps:
\begin{enumerate}
    \item Commute through all terms that do not contribute to the expectation value.
    \item Move the rotation associated to the driver Hamiltonian to the left-hand-side of the observable.
    \item Expand the driver rotation as a sum over shift-operators.
    \item Move the shift operators to the left-hand-side of each term in the sum.
    \item Project each resulting term in the sum onto the identity.
\end{enumerate}

We detail these steps for the single and two body terms separately.

\subsection{Single-body terms}

Starting with the single-body terms, commuting through all the terms that do
not contribute, gives:
\begin{align}
    \label{eq:drive_sb}
    e^{-iH_d \beta} Z^r_k e^{-iH_d \beta} & = e^{-i \beta \sum_{i=1}^{m}\sum_{j=0}^{d_i} \Gamma_{ij} X_{i}^j} Z^r_k e^{i \beta \sum_{i=1}^{m}\sum_{j=0}^{d_i} \Gamma_{ij} X_{i}^j}\\
     & = e^{-i \beta \sum_{j=0}^{d_k} \Gamma_{kj} X_{k}^j} Z^r_k e^{i \beta \sum_{j=0}^{d_k} \Gamma_{kj} X_{k}^j}
\end{align}
Pushing a single $X$ rotation past the $Z^r_k$ term gives:
\begin{align}
    Z^r_k e^{i \beta \Gamma_{kj} X_{k}^j}
    & = Z^r_k \sum_{t=0}^{\infty} \frac{\left(i \beta\right)^t}{t!} X_{k}^{jt}\\
    & = \sum_{t=0}^{\infty} \frac{\left(i \beta\right)^t}{t!} \omega_{d_k+1}^{jrt} X_{k}^{jt} Z^r_k\\
    & = \sum_{t=0}^{\infty} \frac{\left(i \beta \omega_{d_k+1}^{jr} X_{k}^{j}\right)^t}{t!}   Z^r_k\\
    & = e^{i \beta \omega_{d_k+1}^{jr} X_k^j}  Z^r_k,
\end{align}
where we have used $Z^r_kX_k^j = \omega_{d_k+1}^{rj} X_k^j Z^r_k$. Hence
Eq.~\eqref{eq:drive_sb} becomes:
\begin{equation}
    e^{-i \beta \sum_{j=0}^{d_k} \Gamma_{kj} X_{k}^j} Z^r_k e^{i \beta \sum_{j=0}^{d_k} \Gamma_{kj} X_{k}^j} = e^{-i \beta \sum_{j=0}^{d_k} \Gamma_{kj} (1-\omega_{d_k+1}^{jr}) X_{k}^j} Z_k^r
\end{equation}

In order to expand the exponential into a sum over powers of the shift operator
we use the following result for applying functions to $X_k$:

\begin{align}
    f(X_k) & = \sum_{a=0}^{d_k} f(\omega^a) \tilde{P}_{k}(a)\\
    & = \sum_{\alpha = 0}^{d_k}\left( \frac{1}{d_k+1} \sum_{a=0}^{d_k} f(\omega_{d_k+1}^a) \omega_{d_k+1}^{-a \alpha}\right) X^{\alpha}_k,
\end{align}
where $\tilde{P}_{k}(a)$ is the projector onto the $a^\text{th}$ eigenvalue of
$X$. An analogous statement holds for functions of $Z$.

So Eq.~\eqref{eq:drive_sb} can be expressed as the sum:
\begin{equation}
    e^{-iH_d \beta} Z^r_k e^{-iH_d \beta} = \sum_{\alpha = 0}^{d_k}\left( \frac{1}{d_k+1} \sum_{a=0}^{d_k} e^{-i \beta \sum_{j=0}^{d_k} \Gamma_{kj} (1-\omega_{d_k+1}^{jr}) \omega_{d_k+1}^{aj}} \omega_{d_k+1}^{-a \alpha}\right) X^{\alpha}_k Z_k^r.
\end{equation}

The next step is to move the rotation associated with the problem Hamiltonian
past each $X_k^{\alpha}$ in the sum.
Starting with the single-body rotations:
\begin{equation}
    \label{eq:sb_step1}
    e^{i \gamma \sum_{i=1}^m\sum_{j=0}^{d_i} \left(W_i - w_{ij}\right)P_i(j)} X_k^\alpha e^{-i \gamma \sum_{i=1}^m\sum_{j=0}^{d_i} \left(W_i - w_{ij}\right)P_i(j)} = e^{i \gamma \sum_{j=0}^{d_k} \left(W_k - w_{kj}\right)P_k(j)} X_k^\alpha e^{-i \gamma \sum_{j=0}^{d_k} \left(W_k - w_{kj}\right)P_k(j)}
\end{equation}

Consider moving a shift operator past a term generated by a projector acting on
a single computational basis state:
\begin{align}
    e^{i \gamma \ket{k}\bra{k}}X^\alpha & = \left(1 + \sum_{t=1}^\infty \frac{\left(i \gamma\right)^t}{t!} \ket{k}\bra{k}\right)X^\alpha\\
    & = X^\alpha 1 + \sum_{t=1}^\infty \frac{\left(i \gamma\right)^t}{t!} \ket{k}\bra{k-\alpha}\\
    & = X^\alpha \left(1 + \sum_{t=1}^\infty \frac{\left(i \gamma\right)^t}{t!} \ket{k-\alpha}\bra{k-\alpha}\right)\\
    & = X^\alpha e^{i \gamma \ket{k-\alpha}\bra{k-\alpha}}
\end{align}

It follows that Eq.~\eqref{eq:sb_step1} equals:
\begin{equation}
 e^{i \gamma \sum_{j=0}^{d_k} \left(W_k - w_{kj}\right)P_k(j)} X_k^\alpha e^{-i \gamma \sum_{j=0}^{d_k} \left(W_k - w_{kj}\right)P_k(j)} = X_k^\alpha  e^{-i \gamma \sum_{j=0}^{d_k} \left(W_k - w_{kj}\right)\left(P_k(j)-P_k(j-\alpha)\right)}
\end{equation}

Repeating this for the two-body term:
\begin{align}
    & e^{i \gamma \lambda \sum_{i_2=1}^m \sum_{j_1=0}^{d_{k}-1}\sum_{j_2=0}^{d_{i_2}-1} \delta_{((k, j_1),(i_2, j_2))\in E}P_{k}(j_1) P_{i_2}(j_2)} X_k^\alpha e^{-i \gamma \lambda \sum_{i_2=1}^m \sum_{j_1=0}^{d_{k}-1}\sum_{j_2=0}^{d_{i_2}-1} \delta_{((k, j_1),(i_2, j_2))\in E}P_{k}(j_1) P_{i_2}(j_2)}\\ &= X_k^\alpha e^{-i \gamma \lambda \sum_{i_2=1}^m \sum_{j_1=0}^{d_{k}-1}\sum_{j_2=0}^{d_{i_2}-1} \delta_{((k, j_1),(i_2, j_2))\in E}\left(P_{k}(j_1)-P_{k}(j_1-\alpha)\right) P_{i_2}(j_2)} \\
    &= X_k^\alpha \prod_{i_2=1}^m e^{-i \gamma \lambda \sum_{j_1=0}^{d_{k}-1}\sum_{j_2=0}^{d_{i_2}-1} \delta_{((k, j_1),(i_2, j_2))\in E}\left(P_{k}(j_1)-P_{k}(j_1-\alpha)\right) P_{i_2}(j_2)}\\
    &= X_k^\alpha \prod_{i_2=1}^m \sum_{\tau_{i_2}=0}^{d_{i_2}} \frac{1}{d_{i_2}+1} \sum_{t_{i_2}=0}^{d_{i_2}} e^{-i \gamma \lambda \sum_{j_1=0}^{d_{k}-1}\sum_{j_2=0}^{d_{i_2}-1} \delta_{((k, j_1),(i_2, j_2))\in E}\left(P_{k}(j_1)-P_{k}(j_1-\alpha)\right) \delta_{ j_2,t_{i_2}}} \omega_{d_{i_2}+1}^{-t_{i_2} \tau_{i_2}} Z_{i_2}^{\tau_{i_2}}
\end{align}

Since any term not proportional to the identity will contribute zero, and all
$Z_k$ terms excluding $Z_r$ enter through the above expression, we only need to
consider the case $\tau_{i_k} = 0$ in the above expression.

Putting it all together for $\langle Z^r \rangle$ we have:

\begin{multline}
    \langle Z^r_k \rangle = \sum_{\alpha = 0}^{d_k} \bra{+} X^{\alpha}_k \left( \frac{1}{d_k+1} \sum_{a=0}^{d_k} e^{-i \beta \sum_{j=0}^{d_k} \Gamma_{kj} (1-\omega_{d_k+1}^{jr}) \omega_{d_k+1}^{aj}} \omega_{d_k+1}^{-a \alpha}\right)\\e^{-i \gamma \sum_{j=0}^{d_k} \left(W_k - w_{kj}\right)\left(P_k(j)-P_k(j-\alpha)\right)} \\\prod_{i_2=1}^m  \frac{1}{d_{i_2}+1} \sum_{t_{i_2}=0}^{d_{i_2}} e^{-i \gamma \lambda \sum_{j_1=0}^{d_{k}-1}\sum_{j_2=0}^{d_{i_2}-1} \delta_{((k, j_1),(i_2, j_2))\in E}\left(P_{k}(j_1)-P_{k}(j_1-\alpha)\right) \delta_{j_2,t_{i_2}}}  Z_k^r \ket{+}.
\end{multline}
Acting with $X$ onto the adjacent bra and projecting the inner terms onto the
identity gives:
\begin{multline}
    \label{eq:qaoap1_1b}
    \langle Z^r_k \rangle = \sum_{\alpha = 0}^{d_k} \left( \frac{1}{d_k+1} \sum_{a=0}^{d_k} e^{-i \beta \sum_{j=0}^{d_k} \Gamma_{kj} (1-\omega_{d_k+1}^{jr}) \omega_{d_k+1}^{aj}} \omega_{d_k+1}^{-a \alpha}\right) \times\\ \frac{1}{d_k+1}\sum_{u=0}^{d_k+1} \left[ e^{-i \gamma \sum_{j=0}^{d_k} \left(W_k - w_{kj}\right)\left(\delta_{u,j}-\delta_{u,j-\alpha}\right)} \prod_{i_2=1}^m  \frac{1}{d_{i_2}+1} \sum_{t_{i_2}=0}^{d_{i_2}} e^{-i \gamma \lambda \sum_{j_1=0}^{d_{k}-1}\sum_{j_2=0}^{d_{i_2}-1} \delta_{((k, j_1),(i_2, j_2))\in E}\left(\delta_{u, j_1}-\delta_{u, j_1-\alpha}\right) \delta_{ j_2,t_{i_2}}}  \omega_{d_k+1}^{ru} \right].
\end{multline}

\subsection{Two-body terms}
For the two-body terms the process is much the same as the single-body term.
The question is now, how do the rotations associated with the problem
Hamiltonian move past $X_k^\alpha X_l^{\nu}$. There are two new cases to
consider: interactions between the qudits in question and interactions
involving a common third qudit.

Considering first interactions between the qudits:

\begin{multline}
    e^{i \gamma \lambda \sum_{j_1=0}^{d_{k}-1}\sum_{j_2=0}^{d_{l}-1} \delta_{((k, j_1),(l, j_2))\in E}P_{k}(j_1) P_{l}(j_2)} X_k^\alpha X_l^{\nu} e^{-i \gamma \lambda \sum_{j_1=0}^{d_{k}-1}\sum_{j_2=0}^{d_{l}-1} \delta_{((k, j_1),(l, j_2))\in E}P_{k}(j_1) P_{l}(j_2)} \\ = X_k^\alpha X_l^{\nu} e^{-i \gamma \lambda \sum_{j_1=0}^{d_{k}-1}\sum_{j_2=0}^{d_{l}-1} \delta_{((k, j_1),(l, j_2))\in E}\left(P_{k}(j_1) P_{l}(j_2)-P_{k}(j_1-\alpha) P_{l}(j_2-\nu)\right)}
\end{multline}

And for a qudit adjacent to the two body term in question:

\begin{multline}
    e^{i \gamma \lambda \sum_{j_1=0}^{d_{k}-1}\sum_{j_2=0}^{d_{q}-1} \delta_{((k, j_1),(q, j_2))\in E}P_{k}(j_1) P_{q}(j_2)}e^{i \gamma \lambda \sum_{j_3=0}^{d_{l}-1}\sum_{j_4=0}^{d_{q}-1} \delta_{((l, j_3),(q, j_4))\in E}P_{l}(j_3) P_{q}(j_4)} X_k^\alpha X_l^{\nu}\\ \times e^{-i \gamma \lambda \sum_{j_1=0}^{d_{k}-1}\sum_{j_2=0}^{d_{q}-1} \delta_{((k, j_1),(q, j_2))\in E}P_{k}(j_1) P_{q}(j_2)}e^{-i \gamma \lambda \sum_{j_3=0}^{d_{l}-1}\sum_{j_4=0}^{d_{q}-1} \delta_{((l, j_3),(q, j_4))\in E}P_{l}(j_3) P_{q}(j_4)}\\
    = X_k^\alpha X_l^{\nu}  e^{-i \gamma \lambda \sum_{j_1=0}^{d_{k}-1}\sum_{j_2=0}^{d_{q}-1} \delta_{((k, j_1),(q, j_2))\in E}\left(P_{k}(j_1)-P_{k}(j_1-\alpha)\right) P_{q}(j_2)}e^{-i \gamma \lambda \sum_{j_3=0}^{d_{l}-1}\sum_{j_4=0}^{d_{q}-1} \delta_{((l, j_3),(q, j_4))\in E} \left(P_{l}(j_3)-P_{l}(j_3-\nu)\right) P_{q}(j_4)}
\end{multline}
Projecting the above equation onto the identity for the qudit labelled by $q$
gives:
\begin{multline}
    X_k^\alpha X_l^{\nu} \frac{1}{d_q+1}\sum_{t=0}^{d_q} e^{-i \gamma \lambda \left[\sum_{j_2=0}^{d_{q}-1} \delta_{t,j_2}\left(\sum_{j_1=0}^{d_{k}-1} \delta_{((k, j_1),(q, j_2))\in E}\left(P_{k}(j_1)-P_{k}(j_1-\alpha)\right) + \sum_{j_3=0}^{d_{l}-1} \delta_{((l, j_3),(q, j_2))\in E} \left(P_{l}(j_3)-P_{l}(j_3-\nu)\right) \right)\right]}
\end{multline}

Putting it all together:

\begin{multline}
    \langle Z^r_k Z^s_l \rangle =  \sum_{\alpha = 0}^{d_k} \sum_{\nu = 0}^{d_l} \left( \frac{1}{d_k+1} \sum_{a=0}^{d_k} e^{-i \beta \sum_{j=0}^{d_k} \Gamma_{kj} (1-\omega_{d_k+1}^{jr}) \omega_{d_k+1}^{aj}} \omega_{d_k+1}^{-a \alpha}\right)  \left( \frac{1}{d_l+1} \sum_{a=0}^{d_l} e^{-i \beta \sum_{j=0}^{d_k} \Gamma_{lj} (1-\omega_{d_l+1}^{js}) \omega_{d_l+1}^{aj}} \omega_{d_l+1}^{-a \nu}\right)\\ \bra{+} X_k^{\alpha}X_l^{\nu}\\
    e^{-i \gamma \sum_{j=0}^{d_k} \left(W_k - w_{kj}\right)\left(P_k(j)-P_k(j-\alpha)\right)} e^{-i \gamma \sum_{j=0}^{d_l} \left(W_l - w_{lj}\right)\left(P_l(j)-P_l(j-\nu)\right)}\\
    e^{-i \gamma \lambda \sum_{j_1=0}^{d_{k}-1}\sum_{j_2=0}^{d_{l}-1} \delta_{((k, j_1),(l, j_2))\in E}\left(P_{k}(j_1) P_{l}(j_2)-P_{k}(j_1-\alpha) P_{l}(j_2-\nu)\right)}\\
    \prod_{\substack{i_2\\ (k,i_2)\in E \\ (l,i_2) \not \in E\\ i_2 \neq l}}  \frac{1}{d_{i_2}+1} \sum_{t_{i_2}=0}^{d_{i_2}} e^{-i \gamma \lambda \sum_{j_1=0}^{d_{k}-1}\sum_{j_2=0}^{d_{i_2}-1} \delta_{((k, j_1),(i_2, j_2))\in E}\left(P_{k}(j_1)-P_{k}(j_1-\alpha)\right) \delta_{j_2,t_{i_2}}}\\
    \prod_{\substack{i_2\\ (k,i_2) \not \in E \\ (l,i_2) \in E\\ i_2 \neq k}}  \frac{1}{d_{i_2}+1} \sum_{t_{i_2}=0}^{d_{i_2}} e^{-i \gamma \lambda \sum_{j_1=0}^{d_{l}-1}\sum_{j_2=0}^{d_{i_2}-1} \delta_{((l, j_1),(i_2, j_2))\in E}\left(P_{l}(j_1)-P_{l}(j_1-\nu)\right) \delta_{j_2,t_{i_2}}}\\
    \prod_{\substack{q\\ (k,q) \in E \\ (l,q) \in E\\ q \neq k,l}}  \frac{1}{d_q+1}\sum_{t_q=0}^{d_q} e^{-i \gamma \lambda \left[\sum_{j_2=0}^{d_{q}-1} \delta_{t_q,j_2}\left(\sum_{j_1=0}^{d_{k}-1} \delta_{((k, j_1),(q, j_2))\in E}\left(P_{k}(j_1)-P_{k}(j_1-\alpha)\right) + \sum_{j_3=0}^{d_{l}-1} \delta_{((l, j_3),(q, j_2))\in E} \left(P_{l}(j_3)-P_{l}(j_3-\nu)\right) \right)\right]}
    Z_k^rZ_l^s \ket{+}
\end{multline}
Evaluating the above gives:
\begin{multline}
    \label{eq:qaoap1_2b}
    \langle Z^r_k Z^s_l \rangle = \sum_{\alpha = 0}^{d_k} \sum_{\nu = 0}^{d_l} \left( \frac{1}{d_k+1} \sum_{a=0}^{d_k} e^{-i \beta \sum_{j=0}^{d_k} \Gamma_{kj} (1-\omega_{d_k+1}^{jr}) \omega_{d_k+1}^{aj}} \omega_{d_k+1}^{-a \alpha}\right)  \left( \frac{1}{d_l+1} \sum_{a=0}^{d_l} e^{-i \beta \sum_{j=0}^{d_k} \Gamma_{lj} (1-\omega_{d_l+1}^{js}) \omega_{d_l+1}^{aj}} \omega_{d_l+1}^{-a \nu}\right)\\
    \frac{1}{(d_k+1)(d_l+1)}\sum_{u=0}^{d_k}\sum_{v=0}^{d_l} \omega_{d_k+1}^{ru} \omega_{d_l+1}^{sv}\\
    e^{-i \gamma \sum_{j=0}^{d_k} \left(W_k - w_{kj}\right)\left(\delta_{u,j}-\delta_{u,j-\alpha}\right)} e^{-i \gamma \sum_{j=0}^{d_l} \left(W_l - w_{lj}\right)\left(\delta_{v,j}-\delta_{v,j-\nu}\right)}\\
    e^{-i \gamma \lambda \sum_{j_1=0}^{d_{k}-1}\sum_{j_2=0}^{d_{l}-1} \delta_{((k, j_1),(l, j_2))\in E}\left(\delta_{u,j_1} \delta_{v,j_2} - \delta_{u,j_1-\alpha} \delta_{v,j_2-\nu}\right)}\\
    \prod_{\substack{i_2\\ (k,i_2)\in E \\ (l,i_2) \not \in E\\ i_2 \neq l}}  \frac{1}{d_{i_2}+1} \sum_{t_{i_2}=0}^{d_{i_2}} e^{-i \gamma \lambda \sum_{j_1=0}^{d_{k}-1}\sum_{j_2=0}^{d_{i_2}-1} \delta_{((k, j_1),(i_2, j_2))\in E}\left(\delta_{u,j_1}-\delta_{u,j_1-\alpha}\right) \delta_{j_2,t_{i_2}}}\\
    \prod_{\substack{i_2\\ (k,i_2) \not \in E \\ (l,i_2) \in E\\ i_2 \neq k}}  \frac{1}{d_{i_2}+1} \sum_{t_{i_2}=0}^{d_{i_2}} e^{-i \gamma \lambda \sum_{j_1=0}^{d_{l}-1}\sum_{j_2=0}^{d_{i_2}-1} \delta_{((l, j_1),(i_2, j_2))\in E}\left(\delta_{v,j_1}-\delta_{v,j_1-\nu}\right) \delta_{j_2,t_{i_2}}}\\
    \prod_{\substack{q\\ (k,q) \in E \\ (l,q) \in E\\ q \neq k,l}}  \frac{1}{d_q+1}\sum_{t_q=0}^{d_q} e^{-i \gamma \lambda \left[\sum_{j_2=0}^{d_{q}-1} \delta_{t_q,j_2}\left(\sum_{j_1=0}^{d_{k}-1} \delta_{((k, j_1),(q, j_2))\in E}\left(\delta_{u,j_1} - \delta_{u,j_1-\alpha}\right) + \sum_{j_3=0}^{d_{l}-1} \delta_{((l, j_3),(q, j_2))\in E} \left(\delta_{v,j_3}-\delta_{v,j_3-\nu}\right) \right)\right]}
\end{multline}

We use Eq.~\eqref{eq:qaoap1_2b} to evaluate the performance of QAOA $p=1$ for
circulant adjacency matrices. When the driving between qudit-states corresponds
to a ring the driver Hamiltonian is given by:
\begin{equation}
    \label{eq:ring_gamma}
    \Gamma_{ij}^\text{ring} =
    \begin{cases}
        1 \text{ if } j=1 \text{ or } j = d_i\\
        0 \text{ otherwise}.
    \end{cases}
\end{equation}
For the all-to-all case the driver is given by:
\begin{equation}
    \label{eq:all_gamma}
    \Gamma_{ij}^\text{all-to-all} =
    \begin{cases}
        1 \text{ if } j\neq 0 \\
        0 \text{ otherwise}.
    \end{cases}
\end{equation}
The expectation of the problem Hamiltonian can be recovered from
Eq.~\eqref{eq:qaoap1_1b} and Eq.~\eqref{eq:qaoap1_2b} using:
\begin{equation}
    \langle P_k(j)\rangle = \frac{1}{d_k+1} \sum_{q=0}^{d_k}\omega_{d_k+1}^{-jq} \langle Z_k^q \rangle
\end{equation}
and
\begin{equation}
    \langle P_k(a) P_l(b) \rangle = \frac{1}{(d_k+1)(d_l+1)} \sum_{r=0}^{d_k} \sum_{s=0}^{d_l}  \omega_{d_k+1}^{-ar} \omega_{d_l+1}^{-bs}\langle Z_k^rZ_k^s \rangle.
\end{equation}

\section{Circuit implementation details}
\label{app:circuits}

In this appendix we outline how to construct the QAOA circuits for the clique based driver Hamiltonians used in this paper. Recall that each term in the driver acts on disjoint cliques, so can be considered separately. Inspired by adiabatic quantum optimisation \cite{farhi2000, albash2018}, we show how to implement each of the propagator associated without Trotter error. The absence of Trotter error is achieved by applying a unitary that diagonalises the Hamiltonian, applying rotations proportional to both the angle and associated eigenvalue and then undoing the rotation associated with the diagonalisation. A similar idea was applied in \cite{wangrubin2020} based around the Fast-Fermionic-Fourier Transform (FFFT) \cite{babbush2018, aigner2026} for the ring driver and based around $XX+YY$ Hamiltonians for the all-to-all graph. In this work we instead utilise a Discrete Fourier Transform (DFT) which diagonalises both the ring and all-to-all driver (or any other circulant adjacency matrix) with the same circuit. It also does not rely on the radix-2 nature of the FFFT, or the need for fermionic swaps \cite{babbush2018}. 

As mentioned in the main text each driver Hamiltonian can be written as $H_d = -A$, where $A$ corresponds to the adjacency matrix of a ring, star, or all-to-all graph. The initial-state corresponds to the largest eigenvalue of $A$. Each adjacency matrix is normalised by the difference between its largest and smallest eigenvalue, i.e. the propagator associated with the driver Hamiltonian for one clique in the graph is
\begin{equation}
    U_d(\beta) = e^{{\frac{i}{E_\text{max}-E_\text{min}}}A\beta},
\end{equation}
where $E_\text{max}$ and $E_\text{min}$ correspond to the maximum and minimum eigenvalue of $A$ respectively.

For the \emph{Matching Ring} mixer, each non-trivial clique consists of two nodes. In this case the driver Hamiltonian is given by:
\begin{equation}
    H_d^\text{Matching Ring} \propto - \begin{pmatrix}
    0 & 0 & 0 & 0 \\
    0 & 0 & 1 & 1 \\
    0 & 1 & 0 & 1 \\
    0 & 1 & 1 & 0 \\
    \end{pmatrix} = 
    \underbrace{\frac{1}{\sqrt{3}}\begin{pmatrix}
    1 & 0 & 0 & 0 \\
    0 & 1 & 1 & 1 \\
    0 & 1 & e^{2 \pi i/3} & e^{4 \pi i/3} \\
    0 & 1 & e^{4 \pi i/3} & e^{2 \pi i/3} \\
    \end{pmatrix}}_{:=V}
    \underbrace{\begin{pmatrix}
    0 & 0  & 0 & 0 \\
    0 & -2 & 0 & 0 \\
    0 & 0  & 1 & 0 \\
    0 & 0  & 0 & 1 \\
    \end{pmatrix}}_{:=D}
    \frac{1}{\sqrt{3}}
    \begin{pmatrix}
    1 & 0 & 0 & 0 \\
    0 & 1 & 1 & 1 \\
    0 & 1 & e^{2 \pi i/3} & e^{4 \pi i/3} \\
    0 & 1 & e^{4 \pi i/3} & e^{2 \pi i/3} \\
    \end{pmatrix}^{\dagger}
\end{equation} 
Clearly this Hamiltonian does not couple the $\ket{00}$ to any other state. The missing normalisation (the reciprocal of the difference between the largest and smallest eigenvalue) is clearly 3. To implement $U_d$ as a circuit in this case, we use a KAK decomposition \cite{tucci2005} to implement $V$. This is achieved with three CNOT gates. Implementing the diagonal part $D$ consists of parameterised single qubit $Z$ rotations and a $ZZ$ rotation consisting of two CNOT gates. In total 8 CNOT gates are used to implement the rotation. The initial state in this case is an equal superposition of all three valid states (i.e. $\frac{1}{\sqrt{3}} \left(\ket{01}+\ket{10}+\ket{11}\right)$). The circuit for generating the initial state is shown in Fig.~\ref{fig:matching_initial_state}.

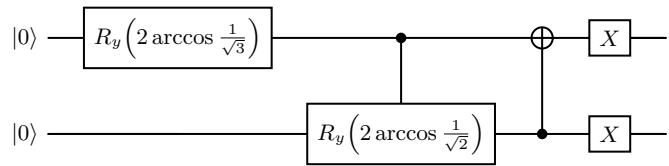
\begin{figure}[H]
\centering
\resizebox{0.5\textwidth}{!}{%
\begin{quantikz}
\lstick{$\ket{0}$} & \gate{R_y\!\left(2\arccos\frac{1}{\sqrt{3}}\right)} & \ctrl{1} & \targ{} & \gate{X} & \qw \\
\lstick{$\ket{0}$} & \qw & \gate{R_y\!\left(2\arccos\frac{1}{\sqrt{2}}\right)} & \ctrl{-1} & \gate{X} & \qw
\end{quantikz}
}
\caption{Quantum circuit for preparing the initial state for each pair of nodes in one edge of a matching.}
\label{fig:matching_initial_state}
\end{figure}

For the Ring and All-to-all driver Hamiltonians, beyond maximal matchings, we utilise one auxiliary qubit per clique in order to one-cold encode the valid states. Therefore, a clique of size $d$ has $d+1$ nodes associated with it. As established all but one node of the clique must be in the solution. If a node is not in the solution, than the associated qubit is assigned the value \mbox{`0'}. If all nodes are in the solution the  auxiliary qubit is assigned \mbox{`1'}, otherwise it is assigned \mbox{`0'}. As a result all valid configuration states correspond to the $d+1$ Hamming weight sector of the Hilbert space associated with the $d+1$ qubits. In practice we use a one-hot encoding by flipping all the qubits, confining the evolution to the single Hamming weight sector. This is achieved by changing the sign of all single qubit $Z$ rotations associated with the problem Hamiltonian and applying a corrective layer of NOT gates at the end of the circuit (as shown in Fig.~\ref{fig:circuit-layer}). 

\begin{figure}[H]
    \centering
    \resizebox{\textwidth}{!}{%
    \begin{tikzpicture}
        \node (circuit) {
            \begin{quantikz}[row sep={0.8cm,between origins}]
                \lstick{$c_1^{(0)}$} & \gate[4]{W-\text{State}} & \qw & \gate{P(w_{c_1^{(0)}}\gamma_k / \lambda)} & \qw  & \ctrl{4}       & \qw & \qw            & \qw & \gate[4]{\text{DFT}^{\dagger}} & \gate{P(\beta_k \tilde{E}_0^{(\abs{c_1})})} & \qw & \gate[4]{\text{DFT}} & \qw & \gate{X} & \qw\\
                \lstick{$c_1^{(1)}$} & \qw                      & \qw & \gate{P(w_{c_1^{(1)}}\gamma_k / \lambda)} & \qw  & \qw            & \qw & \qw            & \qw & \qw                                & \gate{P(\beta_k \tilde{E}_1^{(\abs{c_1})})} & \qw & \qw                  & \qw & \gate{X} & \qw\\
                \lstick{$c_1^{(2)}$} & \qw                      & \qw & \gate{P(w_{c_1^{(2)}}\gamma_k / \lambda)} & \qw  & \qw            & \qw & \ctrl{3}       & \qw & \qw                   & \gate{P(\beta_k \tilde{E}_2^{(\abs{c_1})})} & \qw & \qw                  & \qw & \gate{X} & \qw\\
                \lstick{$c_1^{(a)}$} & \qw & \qw & \qw        & \qw & \qw            & \qw & \qw            & \qw & \qw                   & \gate{P(\beta_k \tilde{E}_3^{(\abs{c_1})})} & \qw & \qw                  & \qw & \gate{X} & \qw\\
                \lstick{$c_2^{(0)}$} & \gate[3]{W-\text{State}} & \qw & \gate{P(w_{c_2^{(0)}}\gamma_k / \lambda)} & \qw & \gate{P(-\gamma_k)} & \qw & \qw            & \qw & \gate[3]{\text{DFT}^{\dagger}} & \gate{P(\beta_k \tilde{E}_0^{(\abs{c_2})})} & \qw & \gate[3]{\text{DFT}} & \qw & \gate{X} & \qw\\
                \lstick{$c_2^{(1)}$} & \qw & \qw & \gate{P(w_{c_2^{(1)}}\gamma_k / \lambda)} & \qw & \qw            & \qw & \gate{P(-\gamma_k)} & \qw & \qw                   & \gate{P(\beta_k \tilde{E}_1^{(\abs{c_2})})} & \qw & \qw                  & \qw& \gate{X} & \qw \\
                \lstick{$c_2^{(a)}$} & \qw & \qw & \qw        & \qw & \qw            & \qw & \qw            & \qw & \qw                   & \gate{P(\beta_k \tilde{E}_2^{(\abs{c_2})})}& \qw & \qw                  & \qw & \gate{X} & \qw \\
            \end{quantikz}
        };
        \draw[draw=prep_colour, fill=prep_colour!10, fill opacity=0.2, rounded corners]
            ($(circuit.north west) + (1 cm, +0.3cm)$)
            rectangle
            ($(circuit.south west) + (3.6cm, 0.3cm)$);
        \node[above, font=\small, text=prep_colour]
            at ($(circuit.north west) + (2.3cm, 0.4 cm)$) {State prep.};
        \draw[draw=prob_colour, fill=prob_colour!10, fill opacity=0.2, rounded corners]
            ($(circuit.north west) + (3.8 cm, +0.3cm)$)
            rectangle
            ($(circuit.south west) + (11.2 cm, 0.3cm)$);
        \node[above, font=\small, text=prob_colour]
            at ($(circuit.north west) + (7.3 cm, 0.4 cm)$) {Problem unitary};
        \draw[draw=driver_colour, fill=driver_colour!10, fill opacity=0.2, rounded corners]
            ($(circuit.north west) + (11.4 cm, +0.3cm)$)
            rectangle
            ($(circuit.south west) + (17.9 cm, 0.3cm)$);
        \node[above, font=\small, text=driver_colour]
            at ($(circuit.north west) + (14.25 cm, 0.4 cm)$) {Driver unitary};
        \draw[draw=corr_colour, fill=corr_colour!10, fill opacity=0.2, rounded corners]
            ($(circuit.north west) + (18.1 cm, +0.3cm)$)
            rectangle
            ($(circuit.south west) + (19.5 cm, 0.3cm)$);
        \node[above, font=\small, text=corr_colour]
            at ($(circuit.north west) + (18.5 cm, 0.4 cm)$) {Corr.};
        \draw[dashed, draw=repeat_colour, fill=repeat_colour!10, fill opacity=0, rounded corners, line width = 2pt]
            ($(circuit.north west) + (3.7 cm, +1.0 cm)$)
            rectangle
            ($(circuit.south west) + (18 cm, -0 cm)$);
        \node[above, font=\small, text=repeat_colour]
            at ($(circuit.south west) + (10.65 cm, -0.6 cm)$) {Repeat $p$ times};
    \end{tikzpicture}
    }
    \caption{\textbf{An illustrative example of a full QAOA circuit.} The quantum circuit shown consists of two registers $c_1$ and $c_2$, representing two cliques in the clique cover. In this case we consider the ring mixer. Both
    registers are initialised in a $W$-state (shaded, left), which corresponds to the ground-state of the driver Hamiltonian. Single-qubit phase ($P$)
    rotations are applied to implement rotations in proportion to the size of the vertex cover. Controlled phase gates $P(\phi)$ couple registers between cliques to enforce penalties between cliques. The DFT and IDFT blocks, implemented through Givens rotations, perform the discrete Fourier transform and
    its inverse on each register. The DFT diagonalises the driver Hamiltonian, such that it can be implemented through diagonal gates. The middle part of the section is repeated $p$ times with the $(\gamma_k, \beta_k)$ changing between repeating circuit blocks. At the end of the circuit a layer of $X$ gates is applied to move from a one-hot to a one-cold encoding.}
    \label{fig:circuit-layer}
\end{figure}
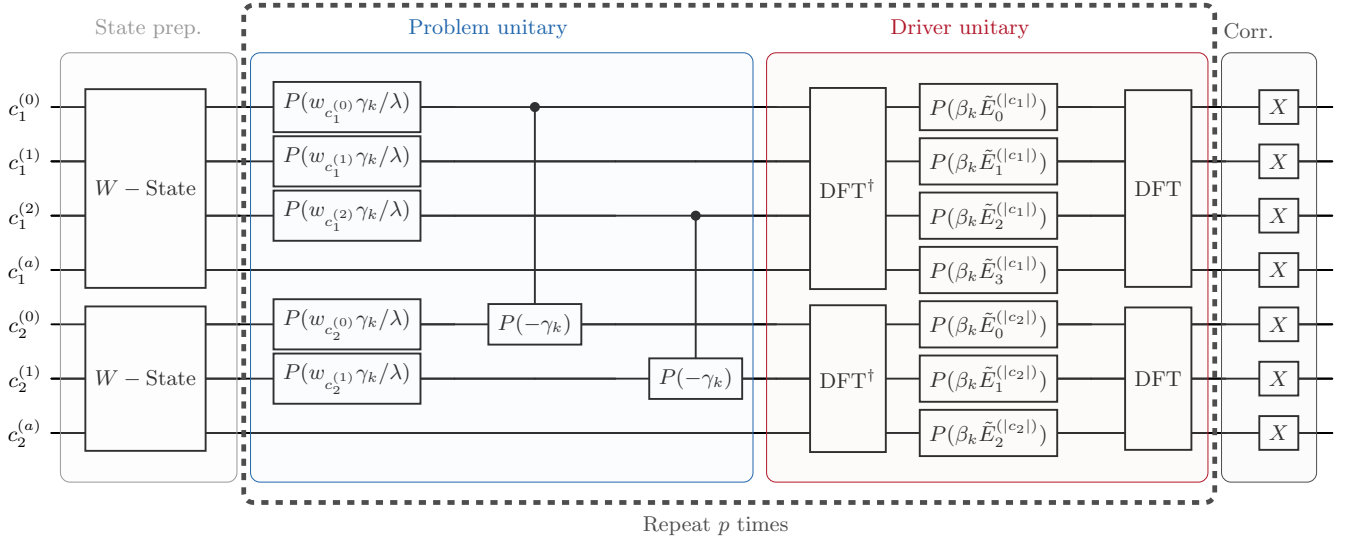

Both the Ring and All-to-all mixers correspond to a circulant matrix and are therefore diagonalised by the Discrete Fourier Transform (DFT). The DFT is implemented through Givens rotations \cite{barkoutsos2018}, and therefore the depth scales in proportion to the size of the clique. For the ring mixer:
\begin{equation}
    A_\text{ring} =\sum_{i=0}^d \ket{i}\bra{i+1}+  \ket{i+1}\bra{i},
\end{equation}
which has eigenvalues $2\cos(2\pi k /(d+1))$ for $k\in\{0,1,\dots,d\}$. The QAOA circuit for this case is illustrated in Fig.~\ref{fig:circuit-layer} where the normalised eigenvalues are denoted by:
\begin{equation}
    \tilde{E}_k^{(d)} = \frac{\cos(2\pi k /(d+1))}{1-\cos(2\pi \lfloor\frac{d+1}{2}\rfloor /(d+1))}.
\end{equation}
For the all-to-all adjacency matrix 
\begin{equation}
    A_\text{all-to-all} =\sum_{j=0}^d \sum_{i>j}^d \ket{i}\bra{j}+  \ket{i}\bra{j}
\end{equation}
the eigenvalues are $d$ corresponding to the equal superposition state and -1 otherwise. The implementation of this driver is shown in Fig.~\ref{fig:circuit-a2a}.

The initial state for both the all-to-all and ring case, the ground-state of the driver Hamiltonian, corresponds to a $W$-state (i.e. an equal superposition of states with Hamming weight 1). It is already known how to implement $W$-states \cite{Cruz_19}. 

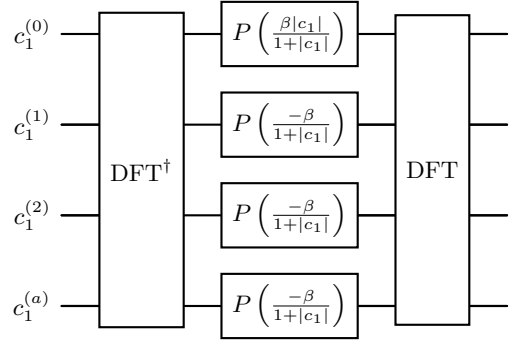
\begin{figure}[H]
    \centering
        \begin{quantikz}[row sep={1.2cm,between origins}]
            \lstick{$c_1^{(0)}$} &  \gate[4]{\text{DFT}^{\dagger}} & \gate{P\left(\frac{\beta \abs{c_1}}{{1+\abs{c_1}}}\right)} & \gate[4]{\text{DFT}} & \qw \\
            \lstick{$c_1^{(1)}$} &  \qw                            & \gate{P\left(\frac{-\beta }{{1+\abs{c_1}}}\right)}      & \qw                  & \qw \\
            \lstick{$c_1^{(2)}$} &  \qw                            & \gate{P\left(\frac{-\beta }{{1+\abs{c_1}}}\right)}      & \qw                  & \qw \\
            \lstick{$c_1^{(a)}$} &  \qw                            & \gate{P\left(\frac{-\beta }{{1+\abs{c_1}}}\right)}      & \qw                  & \qw \\
        \end{quantikz}
    \caption{Trotter-free implementation of the all-to-all driver unitary
    $\exp(i \beta A_\text{all-to-all})$ on a single clique register $c_1$,
    comprising the auxiliary qubit $c_1^{(a)}$. The all-to-all adjacency matrix is
    circulant and therefore diagonalised by the discrete Fourier transform, so
    the driver reduces to single-qubit phase gates conjugated by DFT$^\dagger$
    and DFT.}
    \label{fig:circuit-a2a}
\end{figure}

The star-graph is not circulant and therefore not diagonalised by a DFT. The adjacency matrix on $d+1$ nodes can be written as follows:
\begin{equation}
    A_\text{star} = \sum_{j=0}^{d-1} \ket{j}\bra{d} + \ket{d}\bra{j}
\end{equation}
where $d$ labels the central node. This is a rank-2 matrix with the image spanned by the orthonormal vectors $\ket{d}$ and
\begin{equation}
    \ket{u} = \sum_{j=0}^{d-1} \ket{j}.
\end{equation}
Hence the eigenvectors not in the kernel of $A_\text{star}$ are
$\left(\ket{u}\pm \ket{d}\right)/\sqrt{2}$ with associated eigenvalues $\pm
\sqrt{d}$. Again, we use a one-hot encoding with the all \mbox{`0'} state corresponding to $d$, with the correction pushed to the end of the circuit. Using one auxiliary qubit per clique, we couple $\ket{d}$  to the $\ket{1}$ state of the auxiliary qubit and $\ket{u}$ to the $\ket{0}$ state of the auxiliary qubit. The rotation is then performed on the auxiliary qubit, which is then decoupled from the original qubits. The circuit is shown in Fig.~\ref{fig:circuit-star}. For the initial-state, we prepend the circuit for generating a $W$-state \cite{Cruz_19} by a Hadamard gate acting on the first qubit in the circuit.

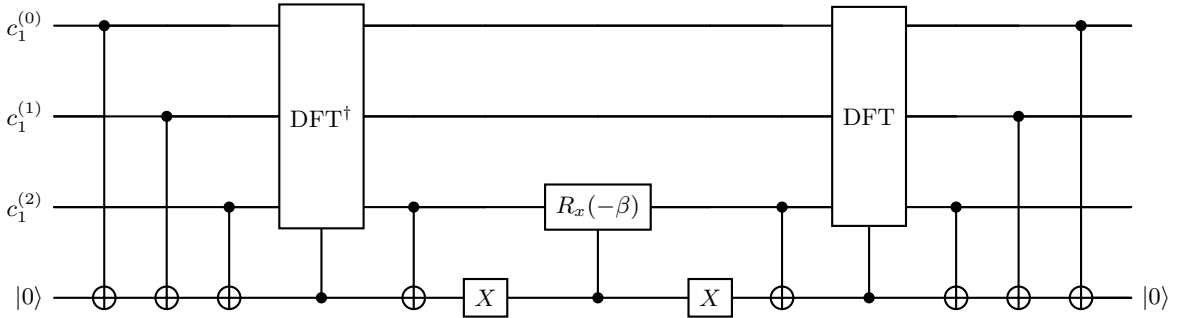
\begin{figure}[H]
    \centering
        \begin{quantikz}[row sep={1.2cm,between origins}]
            \lstick{$c_1^{(0)}$} &  \ctrl{3} & \qw      & \qw      & \gate[3]{\text{DFT}^{\dagger}} & \qw     & \qw      &\qw                 &\qw     & \qw  & \gate[3]{\text{DFT}}    & \qw      & \qw     &  \ctrl{3} & \qw \\
            \lstick{$c_1^{(1)}$} &  \qw      & \ctrl{2} & \qw      & \qw                            & \qw     & \qw      &\qw                 &\qw     & \qw    &\qw   & \qw     & \ctrl{2}  &  \qw     & \qw \\
            \lstick{$c_1^{(2)}$} &  \qw      & \qw      & \ctrl{1} & \qw                            & \ctrl{1}& \qw      &\gate{R_x(-\beta)}  & \qw      & \ctrl{1} & \qw  & \ctrl{1} &  \qw      & \qw    & \qw  \\
            \lstick{$\ket{0}$} &  \targ{}  & \targ{}  & \targ{}  & \ctrl{-1}                      & \targ{} & \gate{X} &\ctrl{-1}           & \gate{X} & \targ{} & \ctrl{-1} & \targ{} & \targ{} &  \targ{} & \qw \rstick{$\ket{0}$} \\
        \end{quantikz}
    \caption{Trotter-free implementation of the star driver unitary
    $\exp(i \beta A_\text{star})$ on a clique register, using one auxiliary qubit
    that is returned to $\ket{0}$. The star graph is not circulant, so the DFT
    diagonalises it only on the subspace excluding the central state $\ket{d}$, and
    the surrounding controlled gates isolate that subspace.}
    \label{fig:circuit-star}
\end{figure}

\section{Monomorphism relaxation}
\label{app:mono}

The induced rule used throughout the main text requires
$\{u,u'\} \in E_1 \Leftrightarrow \{v,v'\} \in E_2$: two candidate pairs may be aligned only
if the interaction is present in both species or absent in both. This is what makes
$\Sthree = 1$ hold by construction, and it is also strict enough to veto biologically real
correspondences. If a protein's interaction profile is genuinely richer in one species than
the other, every mapping that would place it is rejected, not because the orthology is
doubtful but because the surrounding topology is asymmetric.

The monomorphism relaxation weakens the rule to a one-way implication,
$\{u,u'\} \in E_1 \Rightarrow \{v,v'\} \in E_2$, so that the template network may carry edges
the other lacks. The reduction is otherwise unchanged: it remains a hard maximum-weight
clique, and therefore a minimum-weight vertex cover, so the entire pipeline of
Sec.~\ref{sec:qaoa} applies without modification. Only the modular product construction
differs.

\begin{figure*}[t]
    \centering
    \includegraphics[width=\linewidth]{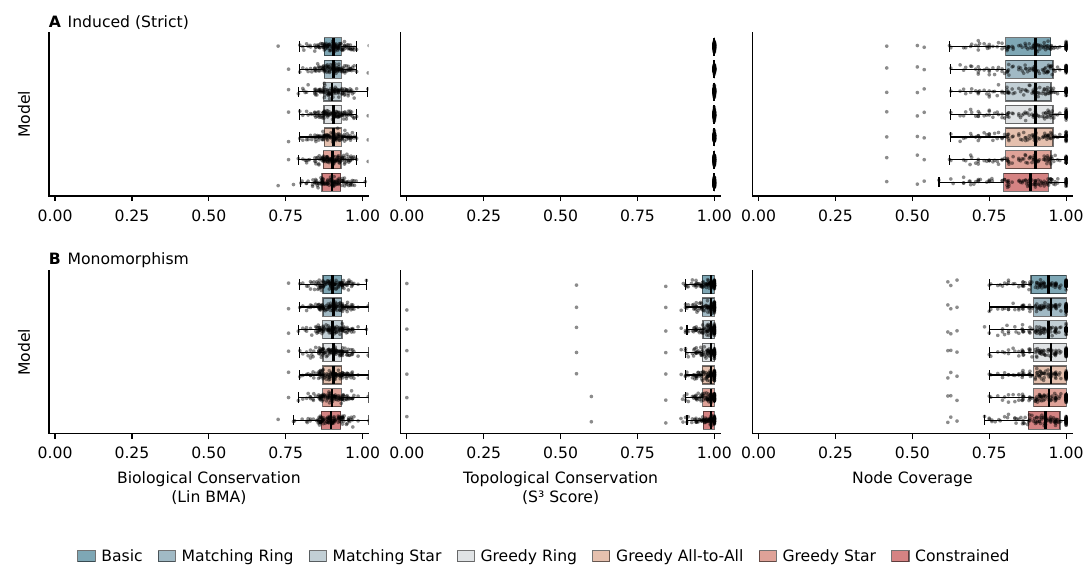}
    \caption{\textbf{Monomorphism relaxation against the strict induced rule under the branch-and-bound
decomposition.} Each panel shows the distribution of alignment quality over the
human--mouse KEGG pathway pairs solved by both formulations, for one metric: biological
conservation (Lin best-match-average GO semantic similarity), topological conservation
($\Sthree$ score), and node coverage. Rows correspond to (A) the strict induced rule and (B) the monomorphism relaxation.
Colours run from blue to red as feasibility moves from the penalty term (\emph{Basic})
into the mixer (\emph{Constrained}).}
    \label{fig:appendix-mono-bnb}
\end{figure*}

\section{Soft-penalty relaxed constraint}
\label{app:relaxed}

A second, independent relaxation acts on the penalty weight rather than on the edge rule.
Sec.~\ref{sec:binary} fixes $\lambda = \max_i w_i + 1$, the smallest value for which no
uncovered edge can ever be profitable. Lowering $\lambda$ below that bound admits a bounded
number of edge violations into the low-energy spectrum: the optimisation is softened, the
sampler is free to pass through infeasible strings, and feasibility is no longer implied by
optimality. This makes it also incompatible with branch and bound decomposition. 

\begin{figure*}[t]
    \centering
    \includegraphics[width=0.9\linewidth]{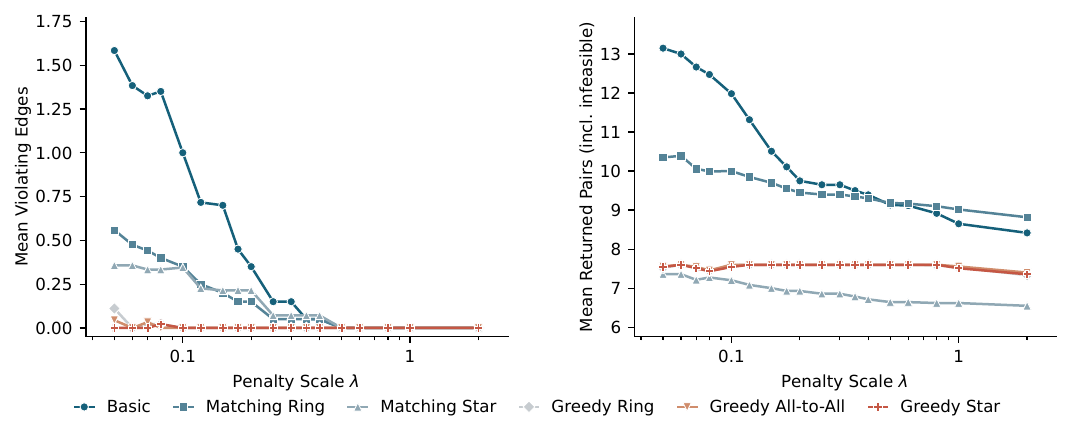}
    \caption{Cost of softening the penalty weight $\lambda$, swept over the small human--mouse
KEGG pathway pairs the solver reliably handles. Each panel plots a mean over pathways
against $\lambda$:  the number of uncovered (violating) edges
left in the returned solution, and the number of returned candidate pairs, counting
infeasible ones. 
Colours run from blue to red as feasibility moves from the penalty term (\emph{Basic}) into
the mixer (\emph{Greedy Star}).}
    \label{fig:penalty-sweep}
\end{figure*}

\begin{figure*}[t]
    \centering
    \includegraphics[width=\linewidth]{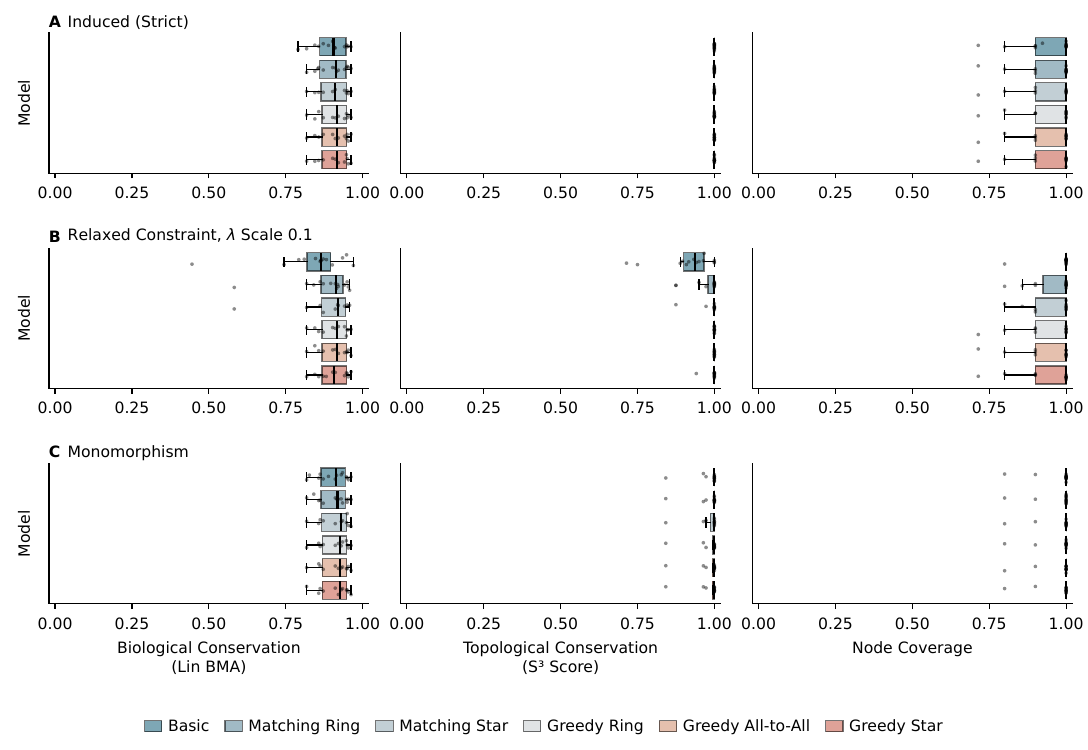}
    \caption{\textbf{Alignment quality of three edge-rule and penalty formulations.} Each panel shows the
distribution of alignment quality over the small human--mouse KEGG pathway pairs for one
metric: biological conservation (Lin best-match-average GO semantic similarity),
topological conservation ($\Sthree$ score), and node coverage. Rows correspond to (A) the
strict induced rule, (B) the monomorphism relaxation, and (C) the soft-penalty relaxation
at $\lambda$ scale $0.1$. Colours run from blue to red as feasibility moves from the
penalty term (\emph{Basic}) into the mixer (\emph{Greedy Star}).}
    \label{fig:appendix-threeway}
\end{figure*}

\FloatBarrier
\section{Datasets details}
\label{app:datasets}

\begin{table}[H]
    \caption{Parameter settings of the synthetic PPI network pairs generated
    with NAPAbench2. The ancestor column gives the two evolutionary scenarios
    with (1) more distantly (2) closely related pairs. Node counts and mean
    degrees are reported separately for the two networks of each pair. Each row
    averages the 10 generated scenarios.}
    \label{tab:napabench}
    \begin{ruledtabular}
    \begin{tabular}{cccccc}
        \multicolumn{2}{c}{Ancestor} & \multicolumn{2}{c}{Network A} & \multicolumn{2}{c}{Network B}\\
        Nodes (Distant) & Nodes (Close) & Nodes & Mean degree & Nodes & Mean degree\\
        \hline
        1 & 3  & 4  & 1.6 & 5  & 1.9\\
        1 & 3 & 4  & 1.7 & 6  & 1.9\\
        2 & 4  & 5  & 1.8 & 5  & 1.8\\
        2 & 4  & 5  & 1.8 & 6  & 2.0\\
        3 & 5  & 6  & 2.0 & 8  & 2.0\\
        3 & 6  & 7  & 2.0 & 10  & 2.2\\
        4 & 6  & 8  & 2.2 & 12  & 2.4\\
        4 & 7  & 9  & 2.3 & 14  & 2.6\\
        4 & 8  & 10  & 2.3 & 16  & 2.6\\
        5 & 9  & 11  & 2.3 & 18  & 2.5\\
        5 & 10  & 12  & 2.4 & 20  & 2.7\\
        5 & 10  & 13  & 2.4 & 24  & 2.9\\
        6 & 11  & 14  & 2.5 & 28  & 2.8\\
        6 & 12  & 15  & 2.4 & 32  & 3.1\\
        7 & 13  & 16  & 2.4 & 36  & 3.2\\
        7 & 14  & 17  & 2.6 & 40  & 3.2\\
        8 & 14  & 18  & 2.7 & 44  & 3.4\\
        8 & 15  & 19  & 2.7 & 48  & 3.4\\
        9 & 16  & 20  & 2.7 & 50  & 3.5\\
    \end{tabular}
    \end{ruledtabular}
\end{table}

\begin{figure}[htb]
    \centering
    \includegraphics[width=0.48\linewidth]{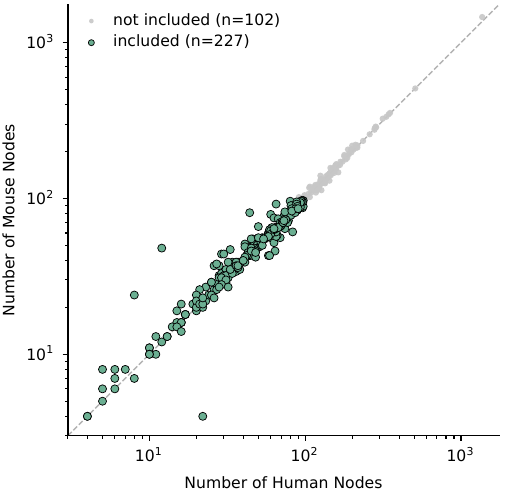}
    \caption{Size of the human--mouse KEGG pathway PPI network pairs. Each point is one
    pathway present in both organisms, placed by the number of nodes in the
    induced human and mouse subgraphs.}
    \label{fig:kegg_dataset}
\end{figure}
\begin{longtable}{llrrrr}
\caption{Human--mouse KEGG pathway pairs used as real-world benchmark
instances, with the size of the induced subgraph in each species.}
\label{tab:keggpathways}\\
\hline
KEGG & Pathway & \multicolumn{2}{c}{Human} & \multicolumn{2}{c}{Mouse}\\
ID & & Nodes & Mean degree & Nodes & Mean degree\\
\hline
\endfirsthead
\hline
KEGG & Pathway & \multicolumn{2}{c}{Human} & \multicolumn{2}{c}{Mouse}\\
ID & & Nodes & Mean degree & Nodes & Mean degree\\
\hline
\endhead
\hline
\endfoot
        00010 & Glycolysis / Gluconeogenesis & 64 & 13.81 & 65 & 13.35\\
        00020 & Citrate cycle (TCA cycle) & 28 & 9.64 & 32 & 9.94\\
        00030 & Pentose phosphate pathway & 29 & 11.38 & 33 & 13.82\\
        00040 & Pentose and glucuronate interconversions & 32 & 5.06 & 33 & 3.33\\
        00051 & Fructose and mannose metabolism & 31 & 10.90 & 35 & 9.83\\
        00052 & Galactose metabolism & 30 & 10.80 & 32 & 11.62\\
        00053 & Ascorbate and aldarate metabolism & 23 & 5.83 & 27 & 3.48\\
        00061 & Fatty acid biosynthesis & 17 & 3.65 & 18 & 3.56\\
        00062 & Fatty acid elongation & 25 & 5.76 & 29 & 6.62\\
        00071 & Fatty acid degradation & 37 & 9.68 & 36 & 10.50\\
        00100 & Steroid biosynthesis & 20 & 5.80 & 20 & 5.80\\
        00120 & Primary bile acid biosynthesis & 17 & 3.53 & 18 & 4.00\\
        00130 & Ubiquinone and other terpenoid-quinone biosynthesis & 11 & 2.36 & 10 & 2.60\\
        00140 & Steroid hormone biosynthesis & 60 & 19.87 & 79 & 12.53\\
        00220 & Arginine biosynthesis & 21 & 7.24 & 20 & 6.40\\
        00232 & Caffeine metabolism & 5 & 4.00 & 6 & 3.67\\
        00240 & Pyrimidine metabolism & 55 & 19.89 & 55 & 20.80\\
        00250 & Alanine, aspartate and glutamate metabolism & 36 & 10.94 & 37 & 11.68\\
        00260 & Glycine, serine and threonine metabolism & 38 & 8.74 & 38 & 8.89\\
        00270 & Cysteine and methionine metabolism & 46 & 10.78 & 50 & 11.56\\
        00280 & Valine, leucine and isoleucine degradation & 46 & 11.83 & 55 & 15.35\\
        00290 & Valine, leucine and isoleucine biosynthesis & 4 & 3.00 & 4 & 3.00\\
        00310 & Lysine degradation & 35 & 3.43 & 38 & 4.42\\
        00330 & Arginine and proline metabolism & 47 & 7.11 & 52 & 8.00\\
        00340 & Histidine metabolism & 21 & 7.14 & 26 & 7.54\\
        00350 & Tyrosine metabolism & 34 & 9.88 & 39 & 10.31\\
        00360 & Phenylalanine metabolism & 16 & 7.38 & 21 & 8.00\\
        00380 & Tryptophan metabolism & 41 & 7.07 & 51 & 10.16\\
        00400 & Phenylalanine, tyrosine and tryptophan biosynthesis & 6 & 5.00 & 8 & 7.00\\
        00410 & beta-Alanine metabolism & 29 & 11.79 & 30 & 12.33\\
        00430 & Taurine and hypotaurine metabolism & 11 & 7.09 & 13 & 8.62\\
        00440 & Phosphonate and phosphinate metabolism & 6 & 3.33 & 6 & 3.33\\
        00450 & Selenocompound metabolism & 16 & 4.88 & 16 & 4.88\\
        00480 & Glutathione metabolism & 50 & 11.28 & 55 & 5.93\\
        00500 & Starch and sucrose metabolism & 31 & 12.00 & 30 & 11.80\\
        00510 & N-Glycan biosynthesis & 49 & 6.12 & 50 & 6.00\\
        00511 & Other glycan degradation & 13 & 2.15 & 13 & 2.15\\
        00512 & Mucin type O-glycan biosynthesis & 29 & 6.55 & 28 & 6.14\\
        00513 & Various types of N-glycan biosynthesis & 37 & 4.27 & 39 & 4.36\\
        00514 & Other types of O-glycan biosynthesis & 22 & 2.09 & 20 & 2.50\\
        00515 & Mannose type O-glycan biosynthesis & 22 & 3.82 & 23 & 3.91\\
        00520 & Amino sugar and nucleotide sugar metabolism & 41 & 6.93 & 49 & 6.61\\
        00524 & Neomycin, kanamycin and gentamicin biosynthesis & 5 & 4.00 & 5 & 4.00\\
        00531 & Glycosaminoglycan degradation & 19 & 3.26 & 21 & 3.52\\
        00532 & Glycosaminoglycan biosynthesis - chondroitin sulfate / dermatan sulfate & 16 & 3.00 & 16 & 2.88\\
        00533 & Glycosaminoglycan biosynthesis - keratan sulfate & 10 & 2.80 & 11 & 3.09\\
        00534 & Glycosaminoglycan biosynthesis - heparan sulfate / heparin & 12 & 1.50 & 12 & 1.50\\
        00561 & Glycerolipid metabolism & 58 & 10.66 & 60 & 10.90\\
        00562 & Inositol phosphate metabolism & 70 & 16.26 & 70 & 15.29\\
        00563 & Glycosylphosphatidylinositol (GPI)-anchor biosynthesis & 23 & 6.00 & 22 & 5.45\\
        00564 & Glycerophospholipid metabolism & 80 & 13.50 & 96 & 13.48\\
        00565 & Ether lipid metabolism & 29 & 9.17 & 44 & 11.00\\
        00590 & Arachidonic acid metabolism & 44 & 12.36 & 81 & 14.10\\
        00591 & Linoleic acid metabolism & 12 & 5.00 & 48 & 8.50\\
        00592 & alpha-Linolenic acid metabolism & 8 & 1.50 & 24 & 7.33\\
        00600 & Sphingolipid metabolism & 47 & 19.96 & 48 & 20.96\\
        00601 & Glycosphingolipid biosynthesis - lacto and neolacto series & 26 & 13.77 & 26 & 13.23\\
        00603 & Glycosphingolipid biosynthesis - globo and isoglobo series & 15 & 5.20 & 16 & 5.62\\
        00604 & Glycosphingolipid biosynthesis - ganglio series & 14 & 5.00 & 15 & 5.87\\
        00620 & Pyruvate metabolism & 36 & 10.00 & 39 & 11.28\\
        00630 & Glyoxylate and dicarboxylate metabolism & 28 & 4.93 & 31 & 5.55\\
        00640 & Propanoate metabolism & 33 & 10.67 & 34 & 11.06\\
        00650 & Butanoate metabolism & 27 & 5.85 & 28 & 8.29\\
        00670 & One carbon pool by folate & 20 & 13.30 & 19 & 12.53\\
        00730 & Thiamine metabolism & 13 & 3.85 & 13 & 5.85\\
        00740 & Riboflavin metabolism & 6 & 4.00 & 7 & 5.14\\
        00750 & Vitamin B6 metabolism & 5 & 4.00 & 8 & 7.00\\
        00760 & Nicotinate and nicotinamide metabolism & 32 & 16.69 & 39 & 18.56\\
        00770 & Pantothenate and CoA biosynthesis & 20 & 4.80 & 20 & 4.70\\
        00785 & Lipoic acid metabolism & 4 & 2.00 & 4 & 2.00\\
        00790 & Folate biosynthesis & 24 & 7.00 & 24 & 6.50\\
        00830 & Retinol metabolism & 65 & 17.57 & 92 & 8.74\\
        00860 & Porphyrin metabolism & 40 & 4.30 & 40 & 2.95\\
        00900 & Terpenoid backbone biosynthesis & 21 & 5.24 & 22 & 5.27\\
        00910 & Nitrogen metabolism & 8 & 2.00 & 7 & 1.43\\
        00920 & Sulfur metabolism & 10 & 3.80 & 11 & 4.00\\
        00970 & Aminoacyl-tRNA biosynthesis & 44 & 3.64 & 43 & 3.86\\
        00980 & Metabolism of xenobiotics by cytochrome P450 & 69 & 15.25 & 56 & 5.93\\
        00982 & Drug metabolism - cytochrome P450 & 64 & 11.84 & 46 & 4.61\\
        00983 & Drug metabolism - other enzymes & 74 & 8.59 & 77 & 5.61\\
        01040 & Biosynthesis of unsaturated fatty acids & 25 & 6.64 & 29 & 6.55\\
        01210 & 2-Oxocarboxylic acid metabolism & 17 & 8.12 & 18 & 9.44\\
        01212 & Fatty acid metabolism & 53 & 8.23 & 56 & 9.39\\
        01230 & Biosynthesis of amino acids & 73 & 9.48 & 77 & 9.43\\
        01521 & EGFR tyrosine kinase inhibitor resistance & 77 & 16.36 & 78 & 15.15\\
        01522 & Endocrine resistance & 92 & 13.96 & 90 & 13.49\\
        01523 & Antifolate resistance & 20 & 5.00 & 24 & 3.67\\
        01524 & Platinum drug resistance & 65 & 6.71 & 64 & 6.53\\
        02010 & ABC transporters & 15 & 2.80 & 19 & 1.37\\
        03008 & Ribosome biogenesis in eukaryotes & 74 & 15.62 & 74 & 15.54\\
        03015 & mRNA surveillance pathway & 87 & 9.68 & 88 & 10.86\\
        03018 & RNA degradation & 64 & 8.75 & 65 & 9.26\\
        03020 & RNA polymerase & 31 & 19.10 & 29 & 18.83\\
        03022 & Basal transcription factors & 44 & 29.41 & 43 & 28.28\\
        03030 & DNA replication & 36 & 11.11 & 34 & 10.47\\
        03050 & Proteasome & 43 & 32.65 & 45 & 35.24\\
        03060 & Protein export & 20 & 6.30 & 22 & 5.82\\
        03320 & PPAR signaling pathway & 52 & 3.15 & 53 & 3.62\\
        03410 & Base excision repair & 21 & 5.71 & 21 & 5.14\\
        03420 & Nucleotide excision repair & 46 & 9.96 & 43 & 9.40\\
        03430 & Mismatch repair & 22 & 7.73 & 21 & 6.86\\
        03440 & Homologous recombination & 36 & 7.78 & 39 & 8.51\\
        03450 & Non-homologous end-joining & 10 & 5.80 & 10 & 5.40\\
        03460 & Fanconi anemia pathway & 51 & 10.90 & 49 & 12.37\\
        04012 & ErbB signaling pathway & 81 & 13.75 & 81 & 13.51\\
        04061 & Viral protein interaction with cytokine and cytokine receptor & 96 & 7.71 & 87 & 6.37\\
        04064 & NF-kappa B signaling pathway & 97 & 8.49 & 89 & 7.57\\
        04070 & Phosphatidylinositol signaling system & 85 & 17.39 & 84 & 17.26\\
        04115 & p53 signaling pathway & 60 & 5.50 & 56 & 5.14\\
        04122 & Sulfur relay system & 7 & 3.14 & 8 & 2.25\\
        04130 & SNARE interactions in vesicular transport & 32 & 17.31 & 31 & 20.52\\
        04136 & Autophagy - other & 31 & 6.84 & 32 & 6.50\\
        04137 & Mitophagy - animal & 63 & 6.13 & 63 & 5.59\\
        04142 & Lysosome biogenesis & 97 & 5.24 & 97 & 5.28\\
        04146 & Peroxisome & 66 & 3.79 & 58 & 3.93\\
        04211 & Longevity regulating pathway & 87 & 10.57 & 90 & 10.84\\
        04213 & Longevity regulating pathway - multiple species & 58 & 9.45 & 57 & 10.11\\
        04215 & Apoptosis - multiple species & 29 & 7.03 & 31 & 7.48\\
        04216 & Ferroptosis & 31 & 1.48 & 30 & 1.60\\
        04260 & Cardiac muscle contraction & 87 & 14.92 & 84 & 14.57\\
        04330 & Notch signaling pathway & 52 & 7.58 & 50 & 8.40\\
        04340 & Hedgehog signaling pathway & 45 & 7.73 & 48 & 7.54\\
        04350 & TGF-beta signaling pathway & 85 & 9.46 & 86 & 9.21\\
        04370 & VEGF signaling pathway & 55 & 12.55 & 56 & 12.61\\
        04392 & Hippo signaling pathway - multiple species & 27 & 4.67 & 26 & 4.46\\
        04512 & ECM-receptor interaction & 87 & 9.08 & 85 & 11.79\\
        04520 & Adherens junction & 69 & 7.33 & 67 & 6.96\\
        04540 & Gap junction & 87 & 10.39 & 84 & 9.69\\
        04610 & Complement and coagulation cascades & 81 & 4.94 & 90 & 6.44\\
        04612 & Antigen processing and presentation & 63 & 9.65 & 75 & 18.93\\
        04614 & Renin-angiotensin system & 16 & 2.25 & 16 & 2.62\\
        04622 & RIG-I-like receptor signaling pathway & 63 & 7.11 & 52 & 7.62\\
        04623 & Cytosolic DNA-sensing pathway & 58 & 8.03 & 43 & 9.81\\
        04640 & Hematopoietic cell lineage & 59 & 3.90 & 62 & 3.48\\
        04657 & IL-17 signaling pathway & 71 & 7.41 & 68 & 7.24\\
        04658 & Th1 and Th2 cell differentiation & 84 & 8.24 & 84 & 7.93\\
        04660 & T cell receptor signaling pathway & 95 & 13.31 & 97 & 13.24\\
        04662 & B cell receptor signaling pathway & 78 & 11.38 & 72 & 11.44\\
        04664 & Fc epsilon RI signaling pathway & 59 & 14.41 & 59 & 14.95\\
        04666 & Fc gamma R-mediated phagosome formation & 89 & 10.04 & 84 & 10.36\\
        04668 & TNF signaling pathway & 96 & 12.50 & 95 & 12.02\\
        04670 & Leukocyte transendothelial migration & 90 & 10.22 & 86 & 11.40\\
        04672 & Intestinal immune network for IgA production & 38 & 3.79 & 35 & 3.31\\
        04710 & Circadian rhythm & 29 & 8.00 & 28 & 8.29\\
        04713 & Circadian entrainment & 91 & 13.87 & 92 & 14.33\\
        04720 & Long-term potentiation & 63 & 13.56 & 62 & 13.71\\
        04721 & Synaptic vesicle cycle & 65 & 9.29 & 65 & 11.17\\
        04727 & GABAergic synapse & 81 & 8.67 & 83 & 11.23\\
        04730 & Long-term depression & 58 & 10.38 & 59 & 10.92\\
        04740 & Olfactory transduction & 34 & 6.76 & 33 & 6.73\\
        04742 & Taste transduction & 48 & 3.71 & 46 & 4.22\\
        04744 & Phototransduction & 25 & 5.76 & 24 & 5.33\\
        04911 & Insulin secretion & 81 & 8.30 & 78 & 8.56\\
        04912 & GnRH signaling pathway & 84 & 11.55 & 84 & 11.48\\
        04913 & Ovarian steroidogenesis & 45 & 4.71 & 55 & 4.65\\
        04914 & Progesterone-mediated oocyte maturation & 93 & 11.51 & 88 & 12.09\\
        04916 & Melanogenesis & 94 & 12.15 & 95 & 12.19\\
        04917 & Prolactin signaling pathway & 61 & 10.36 & 65 & 9.75\\
        04918 & Thyroid hormone synthesis & 65 & 7.11 & 57 & 8.14\\
        04920 & Adipocytokine signaling pathway & 59 & 6.85 & 62 & 6.87\\
        04923 & Regulation of lipolysis in adipocytes & 52 & 7.77 & 52 & 7.38\\
        04924 & Renin secretion & 64 & 7.81 & 62 & 7.39\\
        04925 & Aldosterone synthesis and secretion & 89 & 9.12 & 88 & 9.41\\
        04927 & Cortisol synthesis and secretion & 59 & 6.44 & 63 & 6.89\\
        04929 & GnRH secretion & 60 & 9.10 & 62 & 8.74\\
        04930 & Type II diabetes mellitus & 42 & 7.10 & 43 & 8.84\\
        04933 & AGE-RAGE signaling pathway in diabetic complications & 94 & 8.00 & 96 & 8.19\\
        04940 & Type I diabetes mellitus & 33 & 5.76 & 47 & 17.79\\
        04950 & Maturity onset diabetes of the young & 22 & 2.27 & 4 & 1.00\\
        04960 & Aldosterone-regulated sodium reabsorption & 33 & 7.64 & 35 & 7.83\\
        04961 & Endocrine and other factor-regulated calcium reabsorption & 48 & 5.58 & 42 & 6.33\\
        04962 & Vasopressin-regulated water reabsorption & 41 & 6.34 & 39 & 7.54\\
        04964 & Proximal tubule bicarbonate reclamation & 15 & 6.27 & 15 & 6.00\\
        04966 & Collecting duct acid secretion & 22 & 14.09 & 23 & 18.35\\
        04970 & Salivary secretion & 74 & 8.11 & 64 & 9.09\\
        04971 & Gastric acid secretion & 69 & 8.12 & 66 & 8.58\\
        04972 & Pancreatic secretion & 67 & 4.78 & 74 & 5.35\\
        04973 & Carbohydrate digestion and absorption & 41 & 6.78 & 42 & 6.24\\
        04974 & Protein digestion and absorption & 77 & 4.03 & 70 & 4.14\\
        04975 & Fat digestion and absorption & 26 & 3.00 & 37 & 5.41\\
        04976 & Bile secretion & 59 & 5.36 & 43 & 6.23\\
        04977 & Vitamin digestion and absorption & 10 & 1.60 & 10 & 1.60\\
        04978 & Mineral absorption & 28 & 3.71 & 37 & 3.14\\
        04979 & Cholesterol metabolism & 45 & 5.82 & 46 & 4.83\\
        05030 & Cocaine addiction & 48 & 6.54 & 45 & 7.51\\
        05031 & Amphetamine addiction & 64 & 10.31 & 61 & 11.15\\
        05032 & Morphine addiction & 86 & 10.00 & 81 & 10.07\\
        05033 & Nicotine addiction & 36 & 4.72 & 37 & 6.38\\
        05100 & Bacterial invasion of epithelial cells & 69 & 9.25 & 70 & 8.91\\
        05133 & Pertussis & 70 & 5.20 & 67 & 5.88\\
        05134 & Legionellosis & 50 & 3.64 & 56 & 3.79\\
        05140 & Leishmaniasis & 65 & 6.18 & 61 & 5.87\\
        05142 & Chagas disease & 93 & 8.86 & 94 & 9.06\\
        05143 & African trypanosomiasis & 26 & 2.46 & 23 & 2.61\\
        05144 & Malaria & 32 & 1.94 & 27 & 2.30\\
        05146 & Amoebiasis & 88 & 4.34 & 95 & 4.72\\
        05150 & Staphylococcus aureus infection & 53 & 4.79 & 50 & 5.36\\
        05204 & Chemical carcinogenesis - DNA adducts & 74 & 14.00 & 82 & 5.63\\
        05210 & Colorectal cancer & 82 & 9.88 & 86 & 10.00\\
        05211 & Renal cell carcinoma & 60 & 10.07 & 59 & 9.32\\
        05212 & Pancreatic cancer & 71 & 9.35 & 71 & 8.96\\
        05213 & Endometrial cancer & 58 & 9.41 & 57 & 9.33\\
        05214 & Glioma & 71 & 12.00 & 69 & 11.91\\
        05215 & Prostate cancer & 88 & 12.91 & 92 & 11.87\\
        05216 & Thyroid cancer & 37 & 4.05 & 36 & 3.94\\
        05217 & Basal cell carcinoma & 63 & 9.94 & 63 & 9.81\\
        05218 & Melanoma & 72 & 9.14 & 70 & 8.91\\
        05219 & Bladder cancer & 37 & 6.32 & 37 & 5.73\\
        05220 & Chronic myeloid leukemia & 75 & 9.79 & 73 & 8.99\\
        05221 & Acute myeloid leukemia & 65 & 8.03 & 64 & 7.78\\
        05222 & Small cell lung cancer & 89 & 7.71 & 89 & 8.45\\
        05223 & Non-small cell lung cancer & 67 & 10.66 & 65 & 9.75\\
        05230 & Central carbon metabolism in cancer & 63 & 9.27 & 64 & 9.47\\
        05231 & Choline metabolism in cancer & 75 & 11.39 & 73 & 12.11\\
        05235 & PD-L1 expression and PD-1 checkpoint pathway in cancer & 87 & 11.93 & 86 & 11.19\\
        05310 & Asthma & 16 & 6.50 & 14 & 5.71\\
        05320 & Autoimmune thyroid disease & 40 & 6.60 & 40 & 20.80\\
        05321 & Inflammatory bowel disease & 54 & 4.15 & 55 & 4.00\\
        05322 & Systemic lupus erythematosus & 83 & 8.94 & 61 & 5.44\\
        05323 & Rheumatoid arthritis & 72 & 8.50 & 70 & 10.06\\
        05330 & Allograft rejection & 27 & 8.00 & 38 & 21.68\\
        05332 & Graft-versus-host disease & 30 & 9.13 & 44 & 20.91\\
        05340 & Primary immunodeficiency & 28 & 2.57 & 27 & 2.22\\
        05410 & Hypertrophic cardiomyopathy & 82 & 9.39 & 83 & 9.16\\
        05412 & Arrhythmogenic right ventricular cardiomyopathy & 73 & 9.56 & 72 & 9.92\\
        05414 & Dilated cardiomyopathy & 90 & 9.84 & 86 & 9.79\\
        05416 & Viral myocarditis & 50 & 5.84 & 66 & 13.94\\
\end{longtable}

\FloatBarrier
\end{widetext}
\clearpage
\newpage

\bibliographystyle{apsrev4-2}
\bibliography{bibliography}

\end{document}